\documentclass[final,5p,times,twocolumn]{elsarticle}

\usepackage{graphicx}
\usepackage{amsmath}
\usepackage{amssymb}
\usepackage{multirow}
\usepackage{tikz,pgfplots}

\usepackage{lineno}
\usepackage{xcolor}
\usepackage{verbatim}
\usepackage[acronym,shortcuts]{glossaries}
\newacronym{CART}{CART}{classification and regression tree}
\newacronym{F0}{F0}{fundamental frequency}
\newacronym{FST}{FST}{finite-state transducer}
\newacronym{G2P}{G2P}{grapheme-to-phoneme}
\newacronym{MFCC}{MFCC}{mel-frequency cepstral coefficient}
\newacronym{TTS}{TTS}{text-to-speech synthesis}

\usepackage{xcolor}
\usepackage{colortbl}
\definecolor{NNWaveGen}{RGB}{220,220,220}
\definecolor{VocoderGen}{RGB}{230, 247, 255}
\definecolor{WaveCatGen}{RGB}{204, 255, 204}
\definecolor{WaveCatGen}{RGB}{204, 255, 204}
\definecolor{SpecFilterGen}{RGB}{255, 204, 204}
\definecolor{WaveFilterGen}{RGB}{255, 255, 153}
\definecolor{OthersGen}{RGB}{255, 255, 255}

\usepackage{url}

\usepackage[
    type={CC},
    modifier={by-nc-nd},
    version={4.0},
]{doclicense}

\journal{CSL. Elsevier Published version: \url{https://doi.org/10.1016/j.csl.2020.101114}}

\begin{document}

\begin{frontmatter}


\title{ASVspoof 2019: A large-scale public database of synthesized, converted and replayed speech}

\author[nii]{Xin Wang\corref{cor1}}\ead{wangxin@nii.ac.jp}
\author[nii,edi]{Junichi Yamagishi\corref{cor2}}\ead{jyamagis@nii.ac.jp}

\author[eurecom]{Massimiliano Todisco\corref{cor2}}\ead{Massimiliano.Todisco@eurecom.fr}
\author[eurecom]{H\'ector Delgado\corref{cor2}}\ead{delgado@eurecom.fr}
\author[eurecom]{Andreas Nautsch\corref{cor2}}\ead{andreas.nautsch@eurecom.fr}
\author[eurecom]{Nicholas Evans\corref{cor2}}\ead{evans@eurecom.fr} 

\author[inria]{Md Sahidullah\corref{cor2}}\ead{md.sahidullah@inria.fr}

\author[uef]{Ville Vestman\corref{cor2}}\ead{ville.vestman@uef.fi}
\author[uef]{Tomi Kinnunen\corref{cor2}}\ead{tkinnu@cs.uef.fi}

\author[nec]{Kong Aik Lee\corref{cor2}}\ead{k-lee@ax.jp.nec.com}


\author[aalto]{Lauri Juvela}\ead{lauri.juvela@aalto.fi}
\author[aalto]{Paavo Alku}\ead{paavo.alku@aalto.fi}

\author[sinica]{Yu-Huai Peng}\ead{roland19930601@gmail.com}
\author[sinica]{Hsin-Te Hwang}\ead{hwanght@iis.sinica.edu.tw}
\author[sinica]{Yu Tsao}\ead{yu.tsao@citi.sinica.edu.tw}
\author[sinica]{Hsin-Min Wang}\ead{whm@iis.sinica.edu.tw}

\author[adapt]{S\'{e}bastien Le Maguer}\ead{lemagues@tcd.ie}

\author[google]{Markus Becker}\ead{mabecker@google.com}
\author[google]{Fergus Henderson}\ead{fergus@google.com}
\author[google]{Rob Clark}\ead{rajclark@google.com}
\author[google]{Yu Zhang}\ead{ngyuzh@google.com}
\author[google]{Quan Wang}\ead{quanw@google.com}
\author[google]{Ye Jia}\ead{jiaye@google.com}

\author[hoya]{Kai Onuma}\ead{kai.onuma@readspeaker.com}
\author[hoya]{Koji Mushika}\ead{koji.mushika@readspeaker.com}
\author[hoya]{Takashi Kaneda}\ead{taku.kaneda@readspeaker.com}

\author[iflytek]{Yuan Jiang}\ead{yuanjiang@iflytek.com}
\author[iflytek]{Li-Juan Liu}\ead{ljliu@iflytek.com}

\author[nagoya]{Yi-Chiao Wu}\ead{yichiao.wu@g.sp.m.is.nagoya-u.ac.jp}
\author[nagoya]{Wen-Chin Huang}\ead{wen.chinhuang@g.sp.m.is.nagoya-u.ac.jp}
\author[nagoya]{Tomoki Toda}\ead{tomoki@icts.nagoya-u.ac.jp}

\author[ntt]{Kou Tanaka}\ead{kou.tanaka.ef@hco.ntt.co.jp}
\author[ntt]{Hirokazu Kameoka}\ead{kameoka.hirokazu@lab.ntt.co.jp}

\author[aud]{Ingmar Steiner}\ead{isteiner@audeering.com}

\author[avi]{Driss Matrouf}\ead{driss.matrouf@univ-avignon.fr}
\author[avi]{Jean-Fran\c{c}ois Bonastre}\ead{jean-francois.bonastre@univ-avignon.fr}

\author[edi]{Avashna Govender}\ead{s1689645@staffmail.ed.ac.uk}
\author[edipast]{Srikanth Ronanki}\ead{srikanth.ronanki@ed.ac.uk}

\author[ustc]{Jing-Xuan Zhang}\ead{nosisi@mail.ustc.edu.cn}
\author[ustc]{Zhen-Hua Ling}\ead{zhling@ustc.edu}

\cortext[cor1]{Corresponding authors}
\cortext[cor2]{Equal contribution}

\address[nii]{National Institute of Informatics, 2-1-2 Hitotsubashi, Chiyoda-ku, Tokyo, Japan} 
\address[edi]{Centre for Speech Technology Research, University of Edinburgh, UK}
\address[eurecom]{EURECOM, Campus SophiaTech, 450 Route des Chappes, 06410 Biot, France}
\address[inria]{Universit\'{e} de Lorraine, CNRS, Inria, LORIA, F-54000, Nancy, France}
\address[uef]{University of Eastern Finland, Joensuu campus, L\"{a}nsikatu 15, FI-80110 Joensuu, Finland}
\address[nec]{NEC Corp., 7-1, Shiba 5-chome Minato-ku, Tokyo 108-8001, Japan}
\address[aalto]{Aalto University, Rakentajanaukio 2 C, 00076 Aalto, Finland}
\address[sinica]{Academia Sinica, 128, Sec.\ 2, Academia Road, Nankang, Taipei, Taiwan}
\address[adapt]{Sigmedia, ADAPT Centre, School of Engineering, Trinity College Dublin, Ireland}
\address[google]{Google Inc., 1600 Amphitheatre Parkway, Mountain View, CA 94043, USA}
\address[hoya]{HOYA, Shinjuku Park Tower 35F, 3-7-1 Nishi-Shinjuku, Shinjuku-ku, Tokyo 163-1035 Japan}
\address[iflytek]{iFlytek Research, High-tech Development Zone, No. 666 Wangjiang West Road, Hefei, 230088, China}
\address[nagoya]{Nagoya University, Furo-cho, Chikusa-ku, Nagoya, Aichi 464-8601, Japan}
\address[ntt]{NTT Communication Science Laboratories, 3-1, Morinosato Wakamiya Atsugi-shi, Kanagawa, 243-0198 Japan}
\address[aud]{audEERING GmbH, Friedrichshafener Str.\ 1 82205 Gilching, Germany}
\address[avi]{Avignon University, LIA, 339 Chemin des Meinajari\'{e}s, 84911 Avignon, France}
\address[edipast]{Centre for Speech Technology Research, University of Edinburgh, UK (Currently with Amazon)}
\address[ustc]{University of Science and Technology of China, No.96, JinZhai Road Baohe District,Hefei, Anhui, 230026, China}

\begin{abstract}
\emph{Automatic speaker verification} (ASV) is one of the most natural and convenient means of biometric person recognition. Unfortunately, just like all other biometric systems, ASV is vulnerable to spoofing, also referred to as  ``presentation attacks.'' These vulnerabilities are generally unacceptable and call for spoofing countermeasures or ``presentation attack detection'' systems. In addition to impersonation, ASV systems are vulnerable to replay, speech synthesis, and voice conversion attacks.

The ASVspoof challenge initiative was created to foster research on anti-spoofing and to provide common platforms for the assessment and comparison of spoofing countermeasures. The first edition, ASVspoof 2015, focused upon the study of countermeasures for detecting of \emph{text-to-speech synthesis} (TTS) and \emph{voice conversion} (VC) attacks.  The second edition, ASVspoof 2017, focused instead upon replay spoofing attacks and countermeasures. The ASVspoof 2019 edition is the first to consider all three spoofing attack types within a single challenge.  While they originate from the same source database and same underlying protocol, they are explored in two specific use case scenarios. Spoofing attacks within a \emph{logical access} (LA) scenario are generated with the latest speech synthesis and voice conversion technologies, including state-of-the-art neural acoustic and waveform model techniques. Replay spoofing attacks within a \emph{physical access} (PA) scenario are generated through carefully controlled simulations that support much more revealing analysis than possible previously. Also new to the 2019 edition is the use of the tandem detection cost function metric, which reflects the impact of spoofing and countermeasures on the reliability of a fixed ASV system.

This paper describes the database design, protocol, spoofing attack implementations, and baseline ASV and countermeasure results.
It also describes a human assessment on spoofed data in logical access. It was demonstrated that the spoofing data in the ASVspoof 2019 database have varied degrees of perceived quality and similarity to the target speakers, including spoofed data that cannot be differentiated from bona fide utterances even by human subjects. It is expected that the ASVspoof 2019 database, with its varied coverage of different types of spoofing data, could further foster research on anti-spoofing.
\end{abstract}

\begin{keyword}
Automatic speaker verification \sep countermeasure \sep anti-spoofing \sep presentation attack \sep presentation attack detection \sep Text-to-speech synthesis \sep Voice conversion \sep Replay \sep ASVspoof challenge \sep biometrics \sep media forensics
\newline
\doclicenseThis
\end{keyword}

\end{frontmatter}


\section{Introduction}
\label{S:1}

\emph{Automatic speaker verification} (ASV) is one of the most convenient and natural means of biometric person recognition.  It is perhaps for this reason that the technology is now deployed across an ever-growing array of diverse applications and services, e.g., mobile telephones, smart speakers, and call centers.  

While recent times have seen great strides in the performance of ASV systems, it is now accepted that the technology is vulnerable to manipulation through \emph{spoofing}, also referred to as \emph{presentation attacks}.  There are at least four major classes of spoofing attacks: impersonation, replay, speech synthesis, and voice conversion~\cite{SpringerBookChapter2018}. Impersonation attacks involve one person altering their voice so that it resembles that of another person (the target speaker).  The vulnerability of ASV systems to impersonation remains somewhat uncertain.

Replay attacks are the most straightforward to implement; they can be performed through the recording of a bona fide access attempt, presumably surreptitiously, followed by its presentation to the ASV system.  Given that they are simply recordings of bona fide speech, replay attacks can be highly effective in deceiving ASV systems. Furthermore, both recording and presentation can be performed using inexpensive, consumer-grade devices, meaning that replay attacks can be mounted with ease by the lay person.

This is generally not the case for the two other classes of attacks. The mounting of speech synthesis and voice conversion attacks usually calls for specific know-how and/or a familiarity with sophisticated speech technology.  They are hence generally beyond the capabilities of the lay person. Speech synthesis systems can be used to generate entirely artificial speech signals, whereas voice conversion systems operate on natural speech. With sufficient training data, both speech synthesis and voice conversion technologies can produce high-quality speech signals that mimic the speech of a specific target speaker and are also highly effective in manipulating ASV systems.

The threat of spoofing to biometric technology has been known for some decades~\cite{pellom1999experimental}.  The awareness of this threat spawned research on \emph{anti-spoofing}, namely techniques to distinguish between bona fide and spoofed biometric data.  Solutions are referred to as \emph{spoofing countermeasures} or \emph{presentation attack detection} systems~\cite{ISOpresentationAtack}. While there is a body of earlier work, efforts to develop anti-spoofing solutions for ASV have been spearheaded recently through the community-driven ASVspoof initiative\footnote{\url{https://www.asvspoof.org}}.  Born in 2013~\cite{interspeechSpecialSession2013}, ASVspoof sought to collect and distribute common datasets of bona fide and spoofed speech signals and to establish common platforms for the comparative evaluation of different spoofing countermeasure solutions. The result is a series of bi-annual anti-spoofing challenges whereby participants working with a common database produce scores that reflect the likelihood that speech recordings are bona fide (or spoofed) speech.  Participant scores are then compared by the challenge organizers with ground-truth labels in order to estimate and compare the performance of each anti-spoofing solution. 

The first ASVspoof challenge held in 2015 focused on the development of countermeasures for the detection of speech synthesis and voice conversion spoofing attacks~\cite{Wu-ASVspoof2015}. The second edition, held in 2017, focused instead on the development of countermeasures for the detection of replay spoofing attacks~\cite{Kinnunen2017-assessing}. While ASVspoof 2015 and 2017 showed that there is great potential in distinguishing between bona fide and spoofed speech, the threat of spoofing is very much a moving target.  Speech synthesis and voice conversion technologies have evolved considerably over the last four years, whereas a much more controlled setup is needed to study more thoroughly the threat of replay attacks and the limits of replay spoofing countermeasures.  It is critical that progress on anti-spoofing keeps pace with developments in technologies and techniques that can be used to deceive ASV systems.

\begin{figure*}[tb]
\centering\includegraphics[width=0.9\linewidth]{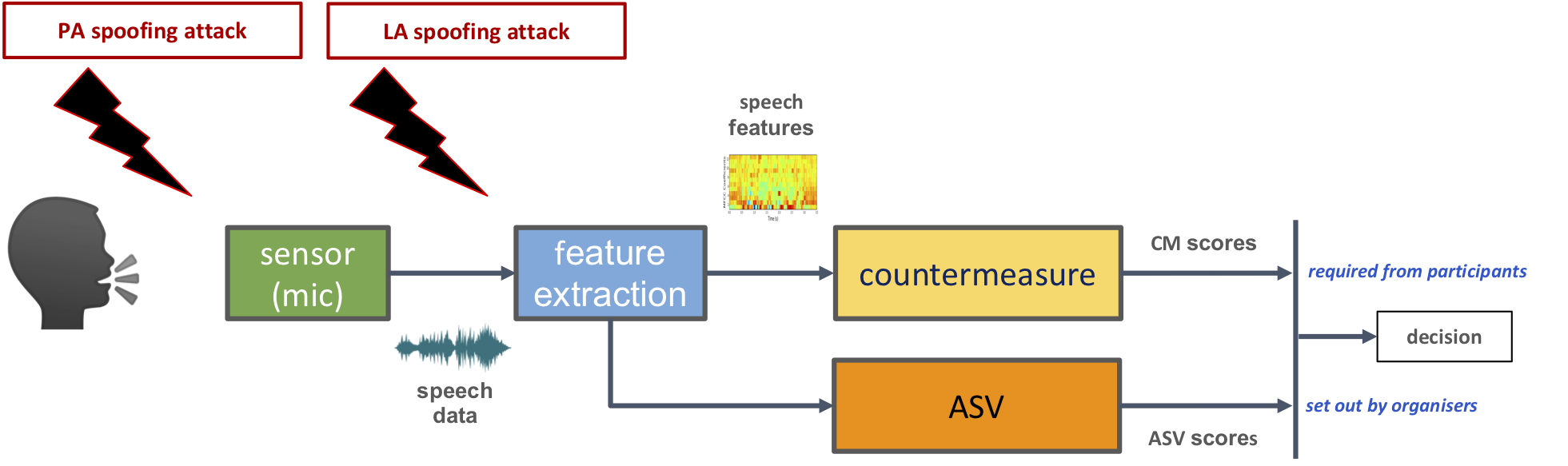}
\caption{LA access and PA access adopted for ASVspoof 2019}
\label{fig:la_pa}
\end{figure*}

The ASVspoof 2019 challenge, accompanied by a new dataset, brings a number of advances compared with previous editions~\cite{ASVspoof19_evalplan,Todisco2019}.  It is the first to consider all three spoofing attack types within a single challenge: speech synthesis, voice conversion, and replay.  While they originate from the same source database and same underlying protocol, they are explored in two specific scenarios illustrated in Figure \ref{fig:la_pa}, namely logical and physical access, with distinct datasets for each scenario.

The ASVspoof 2019 database of bona fide and spoofed speech signals includes synthetic speech and converted voice signals generated with the very latest, state-of-the-art technologies, including Tacotron2~\cite{shen2018natural} and WaveNet~\cite{oord2016wavenet}.  Under well controlled conditions, the best of these algorithms produces spoofed speech that is perceptually indistinguishable from bona fide speech.  ASVspoof 2019 thus aims to determine whether advances in speech synthesis and voice conversion technology pose a greater threat to the reliability of ASV systems or whether, instead, they can be detected reliably with existing countermeasures.

Also new to the 2019 database is carefully controlled simulations of replayed speech that support much more revealing analyses than possible previously. 
\textcolor{black}{The resulting database is suited not only to the study of ASV replay spoofing and countermeasures but also the study of fake audio detection in the case of, \emph{e.g.}\ smart home devices.}

To reflect as best as possible the practical scenario in which the nature of spoofing attacks is never known in advance, a detailed description of the ASVspoof 2019 database was deliberately withheld from participants during the evaluation period.  Now that the evaluation period is complete, a detailed description can finally be released into the public domain. This paper describes (i)~the database design policy, (ii)~the evaluation protocol for both logical and physical access scenarios, (iii)~the techniques and algorithms used in creating the spoofing attacks in each scenario, (iv)~their impact on the reliability of ASV systems, and (v)~two baseline spoofing countermeasures and their performance for each scenario.  Also new in this paper are the results of crowd-sourced human listening tests of spoofing detection. 
The ASVspoof 2019 database is publicly available at \url{https://doi.org/10.7488/ds/2555}.

\section{Database partitions and protocols}
\label{S:2}


In contrast to the ASVspoof 2015 and 2017 databases, the 2019 edition focuses on all three major forms of spoofing attack, namely replay, speech synthesis, and voice conversion. To support such a broader assessment, the ASVspoof 2019 database is designed to reflect two different use case scenarios, namely logical and physical access control.  Also different is the strategy of assessing spoofing and countermeasure performance {\it on ASV}, instead of standalone countermeasure performance.  All of these differences demand an entirely new database design policy.  It is described here.  Training, development, and evaluation partitions are described first.  Protocols for the generation of spoofing attacks in each scenario are described next.  Described last are ASV protocols for assessment with and without spoofing attacks and countermeasures.

\subsection{Database partitions}

\begin{figure*}[tb]
\centering\includegraphics[width=0.9\linewidth]{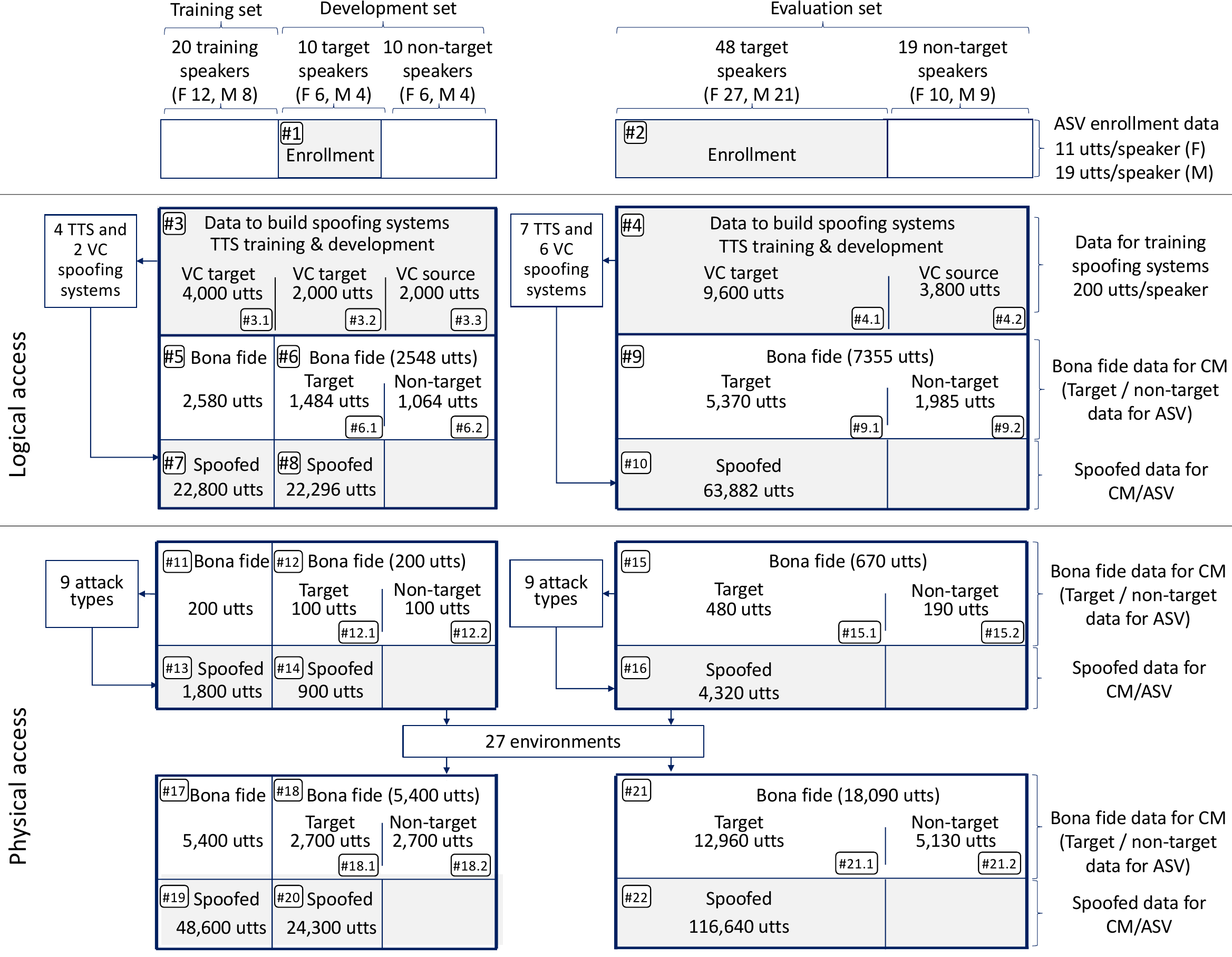}
\caption{Partitions and protocols of ASVspoof 2019 database. Top part shows divisions of training, development, and evaluation sets common to logical access and physical access conditions and numbers of speakers included in each set. Middle part shows database partitions for logical access condition, and bottom part shows those for physical access condition. }
\label{fig:database_protocol}
\end{figure*}

The ASVspoof 2019 database is based upon the 
\emph{Voice Cloning Toolkit} (VCTK) corpus~\cite{vctk}, a multi-speaker English speech database recorded in a hemi-anechoic chamber at a sampling rate of 96~kHz. 
It was created using utterances from 107 speakers (46 male, 61 female) 
that are downsampled 
to 16~kHz at 16~bits-per-sample. 
The set of 
107 speakers 
is partitioned 
into three speaker-disjoint sets
for 
training, development, and evaluation.
This partitioning scheme is illustrated in
the top panel of
Figure~\ref{fig:database_protocol}. 
The training, development, and evaluation sets include 20 training speakers, 10 target and 10 non-target speakers, and 48 target and 19 non-target speakers, respectively. 


Instead of the \emph{equal error rate} (EER) metric, as used for the 2015 and 2019 editions, ASVspoof 2019 adopted a new ASV-centric metric referred to as the tandem detection cost function (t-DCF)~\cite{Kinnunen2018-tDCF}.  Use of the t-DCF means that the ASVspoof 2019 database is designed not for the standalone assessment of spoofing countermeasures (CMs) but their impact on the reliability of an ASV system when subjected to spoofing attacks.  The database partitioning must then support experimentation and evaluation for both ASV and CMs. 
This implies the need for a non-overlapping subset of speaker enrollment data that are needed for both development and evaluation partitions.
After the pruning of low-quality recordings, and while keeping to the same number of enrollment utterances for all speakers of the same gender, there are 
11 enrollment utterances for female speakers and 
19 utterances for male speakers for both development and evaluation partitions. The enrollment data for the development and evaluation speakers is shown in Figure~\ref{fig:database_protocol} with tag \#1 and \#2, respectively.

\subsection{Spoofing protocols}

Two different spoofing protocols are then defined according to one of two use case scenarios: logical and physical access control.  These scenarios are illustrated in Figure \ref{fig:la_pa} and are defined in the following sections.  Specific details for each scenario protocol follow thereafter.

\subsubsection{Scenario definitions}

Logical access (LA) control implies a scenario in which a remote user seeks access to a system or service protected by ASV.  It is assumed in these scenarios that the microphone is not controlled by the authentication system designer and is instead chosen by the user.  An example is a telephone banking service to which remote customers may connect using either their fixed/landline or mobile/cellular telephone.  Physical access (PA) control implies the use of ASV to protect access to a sensitive or secure physical space or facility.  In this scenario, the microphone is controlled by the authentication system designer, not by the user.  An example is the control of access to a goods warehouse using a door-mounted ASV system.

In the LA scenario, it is assumed that spoofing attacks are presented to the ASV system in a worst-case, post-sensor scenario.  Attacks then take the form of synthetic speech or converted voice, which are presented to the ASV system without convolutive acoustic propagation or microphone effects.  In the PA scenario, both bona fide and spoofed speech are assumed to propagate through a physical space prior to acquisition by a fixed system microphone.  In this case, the worst case scenario is assumed to involve replay attacks that involve the presentation of previously recorded bona fide access attempts performed in the same physical space.

\subsubsection{Logical access}
\label{ladb}



The construction of the LA database is illustrated in the middle panel of Figure~\ref{fig:database_protocol}. Unlike the usual ASV databases, it includes data to be used by attacker and defender sides. The application of speech synthesis is referred to as \emph{text-to-speech} (TTS) and \emph{voice conversion} (VC) algorithms from here on. TTS algorithms convert text input to speech, and VC algorithms conduct transformations of input source speech to target speech. These two algorithms can also be combined. 


\paragraph{\textbf{Defender side}}

ASV/CM training data comprises 2,580 bona fide utterances (\#5) and 22,800 spoofed utterances (\#7) generated by using four TTS and two VC algorithms.  The ASV/CM development partition contains 1,484 bona fide target utterances (\#6.1), 1,064 bona fide non-target utterances (\#6.2), and 22,296 spoofed utterances (\#8) generated with the same TTS and VC algorithms.  



The setup for the ASV/CM evaluation set is similar to that for the ASV/CM development set. There are 5,370 bona-fide target utterances (\#9.1) and 1,985 bona fide non-target utterances (\#9.2).  Spoofed data comprises 63,882 utterances (\#10) generated by using 7 TTS and 6 VC spoofing algorithms. 

As in previous editions of ASVspoof, there are \emph{known} and \emph{unknown} spoofing attacks.  Known attacks are those used for generating spoofing TTS/VC attacks in the ASV/CM training and development partitions.  The ASV/CM evaluation partition is created with a mix of 2 known attacks, algorithms also used for generating attacks in the ASV/CM training and development partitions, and 11 unknown attacks.  While the unknown attacks may use similar techniques to the known attacks, the full algorithms are different.  Further details on all spoofing algorithms used in the generation of all spoofing attacks are given in Section~\ref{S:3}

\paragraph{\textbf{Attacker side}}

For building TTS and VC systems that can generate speech similar to a target speaker's characteristics or that can increase the false acceptance ratios of ASV systems, the TTS and VC systems also require an amount of training data. This data is referred to as TTS/VC training data so as to avoid confusion with the training partition for ASV and CM.

The portion with tag \#3 in Figure~\ref{fig:database_protocol} are used to train TTS/VC systems that generate spoofed data in the ASV/CM training and development partitions (\#7 and \#8). For VC systems that require separate source and target training data, the target speaker partitions (\#3.1 and \#3.2) and non-target partition (\#3.3)  are used as the VC target and source data, respectively. The TTS systems use all the utterances in \#3 for training.

The TTS/VC training data with tag \#4 are for TTS/VC systems that synthesize the spoofed data in the ASV/CM evaluation partition (\#10). The sub portions, \#4.1 and \#4.2, are used as VC target and source data for the VC systems, respectively. The TTS systems use all the utterances in \#4 for training. Note that the TTS/VC training data in \#3 and \#4 consist of 200 utterances per speaker, and \textit{these utterances are disjoint from all utterances used for other purposes.} 


Note that the utterances of non-target speakers were converted into the spoofed utterances in a gender-dependent manner for the VC systems. Text inputs randomly sampled from a text corpus in the same domain as the VCTK corpus were used for generating spoofed data from the TTS system.

\subsubsection{Physical access}
\label{padb}

The construction of the PA database is illustrated in the bottom panel of Figure~\ref{fig:database_protocol}.
Its design respects the same high-level partitioning scheme.  There are, however, some differences.  First, since replay attacks stem only from the recording and presentation of bona fide access attempts, there is no need for replay training data.   Next, replay spoofing attacks are generated according to different replay configurations, rather than with different spoofing algorithms.  A \emph{replay configuration} is defined by an acoustic environment (e.g.\ room dimensions) and an attack type (e.g.\ the physical placement of the microphones and loudspeaker in the acoustic environment).  The top row in the lower panel in Figure~\ref{fig:database_protocol} illustrates the protocol for a single acoustic environment for replay spoofing attacks generated according to one of nine different attack types. The figures in the bottom row correspond to the full protocol comprising replay attacks in 27 different acoustic environments.

The training partition contains bona fide and replayed speech.  Bona fide data comprises 200 utterances collected from the 20 training speakers (\#11).  In a single acoustic environment, replay data is generated according to the 9 different attack types, thereby giving 1,800 replay utterances (\#13). The same procedure is applied to the development partition, but only for the 10 target speakers (\#12), thereby giving 900 replay utterances (\#14).  The process is then repeated for the full set of acoustic environments, thereby giving the numbers of utterances illustrated at the bottom of Figure~\ref{fig:database_protocol}. There are 48,600 and 24,300 replay utterances for the training (\#19) and development (\#20) partitions, respectively. The evaluation set is generated in much the same way, only with 48 target and 19 non-target speakers.  These numbers give 4,320 replay utterances for 9 attack types within a single acoustic environment (\#16).  For the full set of 27 acoustic environments, there are then 116,640 replay utterances (\#22). 

The notion of \emph{known} and \emph{unknown} attacks is slightly different in the PA scenario.  Even if all replay attacks in training, development, and evaluation sets are generated according to a fixed set of replay categories (e.g.\ different sized rooms -- see Section~\ref{S:4} for precise details), the impulse responses in each set are different.  In this sense, \emph{all} replay utterances contained in the evaluation set correspond to \emph{unknown} attacks.

\subsection{CM protocols}

Protocols for the CM experimentation are now straightforward.  They include a combination of bona fide and spoofed utterances.  In the context of CM experimentation, both target data and non-target utterances are considered as bona fide.

For the LA scenario, there are 2,580 bona fide utterances (\#5) and 22,800 spoofed utterances (\#7) in the training set, 2,548 bona fide (\#6) and 22,296 spoofed utterances (\#8) in the development set, and 7,355 bona fide (\#9) and 63,882 spoofed utterances (\#10) in the evaluation set.  Equivalent numbers for the PA scenario are illustrated in at the bottom of Figure~\ref{fig:database_protocol}.

\begin{table*}[!t]
\caption{\label{tab:spoofing_technique}Summary of LA spoofing systems. * indicates neural networks. For abbreviations in this table, please refer to Section \ref{S:3}. Note that A04 and A16 use same waveform concatenation TTS algorithm, and A06 and A19 use same VC algorithm.}
\vspace{-5mm}
\begin{center}
\footnotesize
\tabcolsep 3.5pt
    \begin{tabular}{|l|l|l|l|l|l|l|l|l|}
        \hline
            &  Input           & Input processor  & Duration   &  Conversion       & Speaker represent. & Outputs   & Waveform generator & Post process  	\\
        \hline
        A01 & Text             & NLP              & HMM        &  AR RNN*          & VAE*              & MCC, F0          & WaveNet* &     \\
        A02 & Text             & NLP              & HMM        &  AR RNN*          & VAE*              & MCC, F0, BAP     & WORLD    &    \\
        A03 & Text             & NLP              & FF*        &  FF*              & One hot embed.  & MCC, F0, BAP     & WORLD    &  \\
        A04 & Text             & NLP              & -          &  CART             & -                  & MFCC, F0          & Waveform concat. &  \\
        A05 & Speech (human)   & WORLD            & -          &  VAE*            & One hot embed.  & MCC, F0, AP      & WORLD    &  \\
        A06 & Speech (human)   & LPCC/MFCC        & -          &  GMM-UBM          & -                  & LPC              & Spectral filtering + OLA &  \\
        \hline
        A07 & Text             & NLP              & RNN*       &  RNN*             & One hot embed.  & MCC, F0, BA      & WORLD    &  GAN* \\
        A08 & Text             & NLP              & HMM        &  AR RNN*          & One hot embed.  & MCC, F0          & Neural source-filter*  &    \\
        A09 & Text             & NLP              & RNN*       &  RNN*             & One hot embed.  & MCC, F0          & Vocaine  &   \\
        A10 & Text             & CNN+bi-RNN*            & Attention* &  AR RNN + CNN*    & d-vector (RNN)*    & Mel-spectrograms & WaveRNN* &   \\
        A11 & Text             & CNN+bi-RNN*            & Attention* &  AR RNN + CNN*    & d-vector (RNN)*    & Mel-spectrograms & Griffin-Lim \cite{griffin1984signal} &   \\
        A12 & Text             & NLP              & RNN*       &  RNN*             & One hot embed.  & F0+linguistic features & WaveNet* &   \\
        A13 & Speech (TTS)     & WORLD            & DTW        &  Moment matching* & -                  & MCC                       & Waveform filtering &  \\
        A14 & Speech (TTS)     & ASR*             &  -         &  RNN*             & -                  & MCC, F0, BAP              & STRAIGHT &  \\
        A15 & Speech (TTS)     & ASR*             &  -         &  RNN*             & -                  & MCC, F0                   & WaveNet* &  \\
        A16 & Text             & NLP              & -          &  CART             & -                  & MFCC, F0                   & Waveform concat. &  \\
        A17 & Speech (human)   & WORLD            & -          &  VAE*             & One hot embed.  & MCC, F0                   & Waveform filtering &  \\
        A18 & Speech (human)   & MFCC/i-vector    & -          &  Linear           & PLDA               & MFCC                      & MFCC vocoder &  \\
        A19 & Speech (human)   & LPCC/MFCC        & -          &  GMM-UBM          & -                  & LPC                       & Spectral filtering + OLA & \\
        \hline
    \end{tabular}
\end{center}
\end{table*}

\subsection{ASV protocols}

ASV protocols involve some combination of (i)~bona fide target trials, (ii)~non-target (zero-effort impostor) trials, and (iii)~spoofed trials. ASV protocols are specified for the development and evaluation partitions only.  In both cases, and for both LA and PA scenarios, there are protocols for assessing ASV performance in the traditional sense (target and non-target trials) and for the combined assessment of ASV and CMs (target, non-target and spoofed trials).

{The numbers of trials for each partition and scenario are also illustrated in Figure~\ref{fig:database_protocol}.} For the LA scenario development set, there are 1,484 target utterances (\#6.1) and 1,064 non-target utterances (\#6.2) for traditional ASV assessment.  For combined ASV and CM assessment, there are an additional 22,296 spoofed utterances (\#8).  For the evaluation set, there are 5,370 target utterances (\#9.1), 1,985 non-target utterances (\#9.2), and 63,882 spoofed utterances (\#10). Once again, equivalent numbers for the PA scenario are illustrated at the bottom of Figure~\ref{fig:database_protocol}.

\section{Details of LA subset}
\label{S:3}

\subsection{Spoofing algorithms for LA subset}

The LA subset of the new ASVspoof 2019 database is the result of more than six months of intensive work, including contributions from academic and industrial research laboratories. Currently, there are so many different types of TTS and VC systems in the research field and market, and new methods are proposed almost every week at ArXiv.
Therefore, it is difficult to define either the baseline or major methods for TTS and VC. Nevertheless, we considered various types of approaches to the extent that we could, and 17 totally different types of TTS and VC systems were constructed using the TTS/VC training data (\#3 and \#4 in Figure \ref{fig:database_protocol})\footnote{Audio samples are available at  \url{https://nii-yamagishilab.github.io/samples-xin/main-asvspoof2019}.}. Six of them are designated as known spoofing systems, with the other 11 being designated as unknown spoofing systems. Table~\ref{tab:spoofing_technique} summarizes the spoofing systems, which are fundamentally diverse. The known spoofing systems (A01 to A06) include two VC and four TTS systems. Then A07 to A19 (apart from A16 and A19) are the eleven unknown spoofing systems, and A16 and A19 are the known reference systems using the same algorithms as A04 and A06\footnote{Note that A04/A16 also serves as an anchor to our ASVspoof 2015 database. The spoofing method for A04/A16 was used for the ASVspoof 2015 database and was notated S10 in the previous database. S10 was found to be the most difficult spoofing attack. Since the protocols for ASVspoof 2015 and 2019 are totally different, we cannot directly compare results for the two databases, but A04/A16 would imply how CM research advanced rapidly.}. 
The details of each spoofing system are described below.

\paragraph{\textbf{A01}} A neural-network (NN)-based TTS system. This system follows the standard NN-based statistical parametric speech synthesis (SPSS) framework \cite{zen2013statistical} and uses a powerful neural waveform generator called WaveNet \cite{oord2016wavenet}. Attackers may use A01 to attack ASV/CM systems because the WaveNet vocoder can produce high-quality speech that fools CMs.

A01 uses the Festival\_lite  (Flite) \cite{HTSWorkingGroup2014} as the front-end to convert text into a sequence of linguistic features such as phone and pitch-accent labels. It then uses a hidden Markov model (HMM)-based duration model to predict the duration of context-dependent phones. Then, an NN-based acoustic model predicts a sequence of acoustic features, including the Mel-cepstral coefficients (MCCs) of 60 dimensions, interpolated F0, and voicing flags. Finally, the WaveNet vocoder is used to convert the MCCs and F0 into a waveform. 
While the front-end is off-the-shelf, the duration model, the acoustic model, and the WaveNet vocoder were trained using the data for spoofing systems (\#3 in Figure~\ref{fig:database_protocol}). The acoustic model is a combination of a \emph{shallow autoregressive} (SAR) mixture density network \cite{wangARRMDN} and  \emph{variational auto-encoder} (VAE) \cite{doersch2016tutorial}. 
The input linguistic features are processed by two feedforward and two bi-directional \emph{long short-term memory} (LSTM) layers \cite{Graves2008} into 256-dimensional hidden feature vectors. 
The VAE encoder takes both the acoustic and hidden linguistic features as input and produces an utterance-level latent vector with 64 dimensions. The decoder is the SAR, which reconstructs acoustic features given the encoder's output and the hidden linguistic features. The WaveNet vocoder followed the recipe in \cite{wangICASSP2018} but uses a Gaussian distribution to directly model continuous-valued raw waveforms.  

\paragraph{\textbf{A02}} An NN-based TTS system similar to A01 except that the WORLD vocoder \cite{morise2016world} rather than WaveNet is used to generate waveforms. Attackers may use A02 rather than A01 if they cannot collect sufficient data to train the WaveNet vocoder. 
Note that the acoustic model in A02 predicts 25 dimensional band-aperiodicity coefficients (BAPs), in addition to the MCCs and F0s, as the input to the WORLD vocoder. 

\paragraph{\textbf{A03}} An NN-based TTS system similar to A02. Attackers may use A03 because it can be easily built from scratch by using recipes in an open-source TTS toolkit called Merlin \cite{wu2016merlin}. 

A03 used Festival as the front-end to convert input text into a linguistic specification. This linguistic specification is formatted to match HTS style labels \cite{HTSLabeFormat} with state-level alignment. These labels are converted to binary vectors based on HTS-style questions \cite{wu2016merlin}. One-hot-vector-based speaker code is added to the labels. 60-dimensional MCCs, 25-dimensional BAPs and logarithmic F0 were extracted at 5 ms frame intervals using the WORLD vocoder \cite{morise2016world}. 
Using the acoustic features and linguistic vectors, a duration model and an acoustic model were trained. Both models comprise 6 feed-forward hidden layers; each hidden layer has 1,024 hyperbolic tangent units.
Note that the predictions of the duration model are used together with the linguistic vectors to train the acoustic model. 

\paragraph{\textbf{A04}} A waveform concatenation TTS system based on the MaryTTS platform \cite{Schroeder/etal:2011:IS}\footnote{\url{https://github.com/marytts/}} using the voice　building plugin (v5.4) \cite{Steiner/LeMaguer:2018:LREC}.
Attackers may use A04 because the waveform concatenation method preserves the short-term acoustic features of natural speech, and the waveforms generated from A04 may be difficult to detect for CMs. 

The waveform generation in A04 is performed using standard diphone unit selection with target and join cost coefficients. The target cost includes phonological and prosodic features of each unit and its context, while the join cost is calculated from the \acp{MFCC}, log F0, and log F0 delta at the unit boundaries. During concatenation, contiguous diphones are assembled into larger, non-uniform units, while non-contiguous diphones are joined by linearly interpolating one pitch period. Prosodic target features are predicted using \acp{CART} for F0 and duration, while \ac{G2P} conversion uses \acp{FST} created from a pronunciation dictionary adapted from CMUdict.

\paragraph{\textbf{A05}} An NN-based VC system that uses a VAE \cite{hsu2016voice} as the VC conversion model. Attackers may use A05 because it can be trained without using parallel speech data and is more straightforward to build than conventional VC systems. 

The VAE makes it possible to achieve non-parallel VC based on an unsupervised factorization of speech spectral features via autoencoding. Conversion was carried out by encoding the input spectral features into speaker independent latent vectors and decoding them into converted spectral features with the desired target speaker representation. Specifically, a cross-domain VAE \cite{huang2018voice} was trained to decode and encode mel-cepstral coefficients (MCCs) and spectral envelopes. Each speaker has a 128-dimensional trainable speaker embedding vector that is randomly
initialized and optimized jointly with the cross-domain VAE to minimize reconstruction loss.  
During conversion, the MCCs are converted by the trained VAE given the target speaker vector. The converted MCCs, linearly transformed F0, and unmodified aperiodicity are fed to the WORLD vocoder for waveform generation. 

\paragraph{\textbf{A06}} A transfer-function-based VC system \cite{1660175}. Attackers may use A06 if they intend to use a VC technique specifically designed to increase the impostor acceptance rate of ASV.

The principle behind A06 is to analyze the input voice signal following a source-filter model and replace the filters of the input signal by their corresponding of the targeted speaker. After the modification of the filters, the signal is re-synthesized using the original residual signals and the new filters through a classical overlap-add technique.

Two parallel sets of acoustic features are used: MFCC features with mean and variance normalization and the linear prediction cepstral coefficient (LPCC) features which are used to filter the residuals to obtain speech signals. 
The first step is to train an UBM-GMM with 2048 Gaussians on the LPCC-MFCC feature. Given the trained UBM-GMM, we can estimate for each frame (20ms with shift of 10ms) the  posteriori probability of a Gaussian in the UBM-GMM. For each target speaker ${tar}$ and for each Gaussian $g$ in the GMM-UBM, we estimate the mean of all frames belonging to that Gaussian and to that speaker in the LPCC domain, which is denoted as $m_{g,\text{lpcc}}^{\text{tar}}$. For each frame $x$ of the input waveform from a source speaker, we replace its LPCC coefficients by:
$\Sigma_{g=1}^{M}p(g|x)m_{g,lpcc}^{\text{tar}}$,
where $p(g|x)$ is the posteriori probability of Gaussian $g$ given the frame $x$ and is estimated using the LPCC-MFCC UBM-GMM. 

Note that $m_{g,lpcc}^{\text{tar}}$ is estimated if and only if there is enough data belonging to speaker $tar$ and to Gaussian $g$. Otherwise, it is not used for the transformation. 
The conversion is conducted only on speech frames detected by an voice activity detection (VAD) module. Non speech frames are not changed.

\paragraph{\textbf{A07}} An NN-based TTS system. 
Attackers may use A07 if they intend to leverage the GAN-based post-filter, with the hope that the GAN filter may mask differences between the generated speech waveform and natural speech waveform. 

A07 generates speech in two steps. The first step, which is similar to {A03}, converts the input text into a synthetic waveform using the pipeline of a text analyzer, an acoustic model, a duration model, and the WORLD vector. The acoustic feature vector generated by the acoustic model contains the 60-dimensional Mel-cepstrum, 1-dimensional F0, 5-dimensional aperiodic component, and 1-dimensional unvoiced/voiced (U/V) information. 
The acoustic model is an NN with three 256-dimensional LSTM layers and a linear output layer. The duration model is an NN with a LSTM layer of 64 dimensions and a linear output layer. The alignment for model training was extracted using the Merlin toolkit \cite{wu2016merlin}.

After the waveform is synthesized using the WORLD vocoder, WaveCycleGAN2 \cite{wavecyclegan}, a time-domain neural postfilter, is used to transform the output waveform of the WORLD vocoder into a natural-sounding waveform. In other words, both the input and output of WaveCycleGAN2 are raw waveforms. 
WaveCycleGAN2 was trained using the GAN criterion with four 
discriminators (waveform, Mel-spectrum, MCC, and phase spectrum discriminators). The generator consists of a linear projection layer followed by six stacked ResBlock with 64 units and a linear projection output layer. 

\paragraph{\textbf{A08}} An NN-based TTS system similar to {A01}. 
However, A08 uses a neural-source-filter waveform model \cite{8682298}, which is much faster than WaveNet. Attackers may consider this system if they want to generate fake speech at a high speed.
Another difference from A01 is that A08 replaces the VAE with one-hot speaker vectors. 

\paragraph{\textbf{A09}} An NN-based SPSS TTS system \cite{zen2016fast}. 
Attackers may try  A09 because it is designed for real-time TTS on mobile devices to reduce computation load and disk footprint.

A09 uses two LSTM-based acoustic models and a vocoder called Vocaine \cite{agiomyrgiannakis2015vocaine}.  The first LSTM model is a duration model that takes as input a vector describing the linguistic characteristics for each phoneme in a similar way to the above Merlin toolkit and predicts the number of acoustic frames required for that phone. Given the predicted number of acoustic frames and a very similar linguistic input, the second LSTM model predicts the acoustic features that are used by Vocaine to generate waveforms. The primary differences between this TTS system and that described in \cite{zen2016fast} is that this TTS system uses wider LSTM layers and one-hot speaker for the two LSTM models. In addition to VCTK speech data, proprietary voice data was mixed in order to stabilize and generalize multi-speaker modeling.

\paragraph{\textbf{A10}} An end-to-end NN-based TTS system \cite{jia2018transfer} that applies transfer learning from speaker verification to a neural TTS system called Tacotron 2 \cite{shen2018natural}. Attackers may implement A10 since it is reported that synthetic speech produced by this system has high naturalness and good similarity to target speakers perceptually. 

The base system, Tacotron 2, is composed of two components: a sequence-to-sequence model that generates Mel-spectrograms on the basis of the input text or phoneme sequences and a neural vocoder for converting Mel-spectrograms to a waveform. On top of that, a speaker encoder separately trained for a speaker verification task \cite{wan2018generalized} is used for encoding a few seconds of audio clips for a target speaker into a fixed dimensional speaker embedding, which is used as a condition in the base Tacotron 2 model. 
For the neural vocoder, WaveRNN  \cite{DBLP:journals/corr/abs-1802-08435} was used. 

\paragraph{\textbf{A11}} A neural TTS system that is the same as A10 except that A11 uses the Griffin-Lim algorithm \cite{GriffinL84_SpecInversion_TASSP} to generate waveforms. 
Attackers may consider A11 instead of A10 because the Griffin-Lim algorithm requires no model training and is faster in waveform generation than the WaveRNN in A10.

\paragraph{\textbf{A12}} A neural TTS system based on the AR WaveNet \cite{oord2016wavenet}. Although this model is not real-time, attackers may use it because it produces high-quality waveforms. Phone duration and F0 were first predicted by the above TTS system {A09}; they were then used as input features to an AR WaveNet with a conditioning stack to receive the linguistic features. 

\paragraph{\textbf{A13}} A combined NN-based VC and TTS system. 
Attackers may use A13 because it not only combines TTS and VC but also includes a waveform modification approach to produce output waveforms. Speech generated from such a system may be difficult to detect by CMs.

The TTS system in A13 is a product called VoiceText Web-API, which is publicly available for non-commercial uses\footnote{\url{http://dws2.voicetext.jp/tomcat/demonstration/top.html}}. This TTS system uses a unit selection approach and was built with 40 hours of a single speaker's recordings. Using this TTS system, a parallel database of the TTS voice and bona-fide speech of target speakers in the LA subset was constructed, and a conventional VC model was trained. 
The VC model consists of a highway network and a 3-layered feedforward network, where its inputs are MCCs of the source speaker and outputs are those of the target speaker. This network was trained by using a moment-matching-based loss function \cite{li2015generative}. 
Direct waveform modification was used to generate the output waveform. Given the converted MCCs, a differential spectrum filter was estimated, and the TTS voice was filtered and converted into the waveform of a target speaker \cite{kobayashi2018intra}. 

\paragraph{\textbf{A14}} Another combined VC and TTS system. 
Attackers may consider A14 because the VC part in A14 was built following the best VC system in Voice Conversion Challenge 2018 \cite{ljliu2018wav}.

For A14, a commercial English TTS synthesis engine was first adopted to produce the voice of a female speaker for input text. Then, the TTS voice was used as a source speaker of a VC model and was converted to each of the target speakers. In the VC part, bottleneck features were first extracted from the waveforms of the
source speaker via an automatic speech recognition (ASR) model in order to acquire linguistic-related embedding vectors automatically. The ASR model was trained using an in-house dataset that contains around 3,000 hours of speech. Then, an LSTM-based acoustic model was constructed for each target speaker that predicted acoustic features from the bottleneck features. The acoustic features used are 41-dimensional MCCs, F0, and 5-dimensional BAPs, all of which were extracted 
using STRAIGHT \cite{Kawahara1999Restructuring}. 
For waveform reconstruction, the STRAIGHT vocoder was used.

\paragraph{\textbf{A15}} Another combined VC and TTS system similar to A14. However, A15 uses speaker-dependent WaveNet vocoders rather than the STRAIGHT vocoder to generate waveforms. 
Attackers may use A15 if they intend to improve the speech quality and increase the chance to fool a CM.
These WaveNet vocoders ($\mu$-law companding and 10-bit quantization) were trained by pre-training them using an in-house multi-speaker corpus and fine-tuning it using the speech of target speakers \cite{ljliu2018wav}.

\paragraph{\textbf{A16}} A waveform concatenation TTS system that uses the same algorithm as A04. However, A16 was built given the VCTK data for training spoofing systems in the LA evaluation set (\#4 in Figure~\ref{fig:database_protocol}). 

\paragraph{\textbf{A17}} A NN-based VC system that uses the same VAE-based framework as {A05}. However, rather than using the WORLD vocoder, \textbf{A17} uses a generalized direct waveform modification method \cite{2019arXiv190711898H}\cite{kobayashi2018intra} for waveform generation. Attackers may consider A17 because this method was judged to have the highest spoofing capability in Voice Conversion Challenge 2018 \cite{Kinnunen2018}. 

In this method, an F0 transformed residual signal is first generated from the input speech waveform by applying WSOLA \cite{verhelst1993overlap}, inverse filtering, and resampling processes.  Then, the converted speech waveform is generated by filtering the F0 transformed residual signal with a synthesis filter designed on the basis of the converted Mel-cepstrum. Here, the \emph{mel-log spectrum approximation} (MLSA) filter \cite{tokuda1994mel} was chosen for both inverse and synthesis filtering.

\paragraph{\textbf{A18}} A non-parallel VC system \cite{7953215} inspired by the standard \emph{i-vector} framework \cite{Dehak2011-frontendFA,Kenny2012-small} used in text-independent ASV. Attackers may use A18 because it is based on a transfer learning method where a traditional ASV system (i-vector PLDA, with MFCC inputs) is first trained to optimize ASV. This i-vector PLDA system is then directly used to define the voice conversion function in a regression setting. Attackers may also use A18 because it does not require parallel training data, either. 

The system uses a vocoder that generates speech from MFCCs that are originally not designed for VC. This system proceeds in five steps:
\begin{enumerate}
    \item The first off-line step is to learn an i-vector extractor, that is, a sequence-to-vector encoder, in order to represent utterances of different lengths as fixed-sized i-vectors \cite{Dehak2011-frontendFA,Kenny2012-small}. 
    \item The second offline step is to learn a subspace in the i-vector space that best discriminates speakers and acquire a speaker factor extractor via a probabilistic linear discriminant (PLDA) model \cite{DBLP:conf/iccv/PrinceE07}\cite{DBLP:conf/odyssey/Kenny10}.
    \item The third step is VC training. Given some training i-vectors of the source and target speakers, a linear model that predicts a target speaker i-vector, given a (new) source speaker i-vector, is trained.
    \item The fourth step is voice conversion. From a given new source speaker utterance, its MFCCs and i-vector are extracted, and the MFCCs are converted given the target speaker's predicted i-vector.
    \item The fifth step, the vocoder generates the speech waveform given the transformed target MFCCs and a linearly transformed F0. 
\end{enumerate}
This VC system is the same as the one described in \cite{7953215} except for the vocoder and a few updates in the data engineering methods. The mechanism of the new vocoder is similar to the transfer-function-based approach in {A06}. However, the vocoder here  modifies the source residual signal to match the given target MFCC spectral envelope and pitch. The residual signal is first pitch-shifted to match the modified F0. Pitch-shifting is computed by extracting two-period pulses from the original excitation signal, stretching the pulses to match the modified F0, and generating the excitation signal using PSOLA. Finally, the modified excitation is fed to the synthesis filter computed using the target all-pole envelope derived from the converted MFCCs.

\paragraph{\textbf{A19}} A transfer-function-based VC system using the same algorithm as A06. However, A19 was built given the VCTK data for training spoofing systems in the LA evaluation set (\#4 in Figure~\ref{fig:database_protocol}).

\noindent 

\subsection{Visualization of LA subset}

In this section, we visualize the spoofed and bona-fide data in a 2D space. First, we extracted a 512-dimensional x-vector \cite{snyder2018x} for each of the utterances in the LA subset, which contains 121,461 utterances in total. The neural-network-based x-vector extractor was trained using the Kaldi toolkit and the VoxCeleb recipe\footnote{\url{https://github.com/kaldi-asr/kaldi/tree/master/egs/voxceleb/v2}}. 
Then, the raw x-vectors were whitened given the mean and standard deviation vectors calculated for each speaker, and the whitened x-vectors were further normalized to have a unit length. After that, the processed x-vectors were transformed using \emph{within-class covariance normalization} (WCCN) \cite{wcnn}, where the 19 attacking systems and the bona fide data were treated as 20 different classes. Finally, 
the tree-based t-SNE algorithm \cite{maaten2008visualizing} with a perplexity value of 40 was used to reduce the dimensions of the processed x-vectors. 

\begin{figure}[!t]
	\centering
	\includegraphics[width=1.0\linewidth]{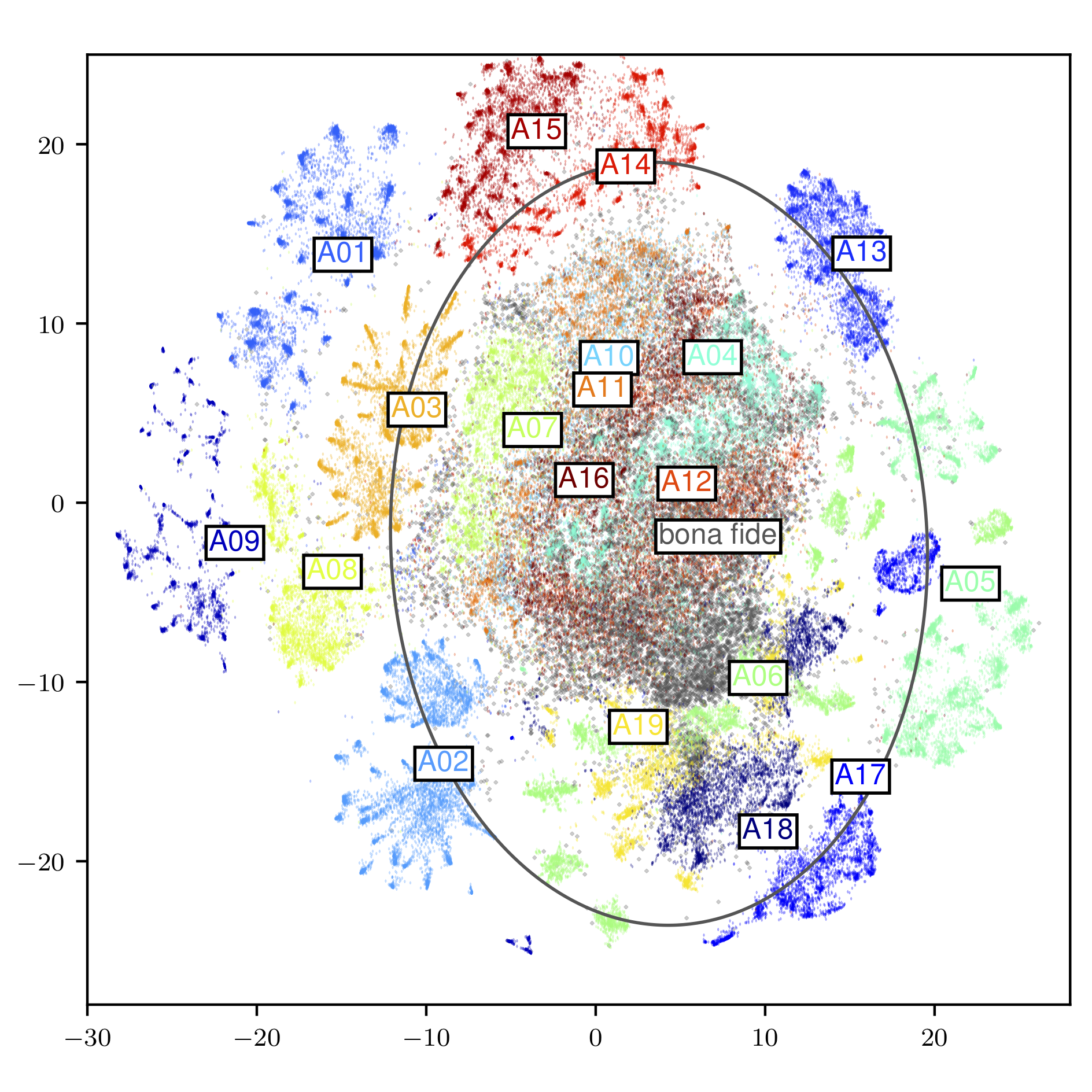}
	\caption{Visualization of bona fide and spoofed speech data of ASVspoof 2019 LA subset. Black circle denotes range of bona fide data within mean $\pm 3$ standard deviation.} 
	\label{fig:visualization_xvector}
	\centering
	\includegraphics[width=0.9\linewidth]{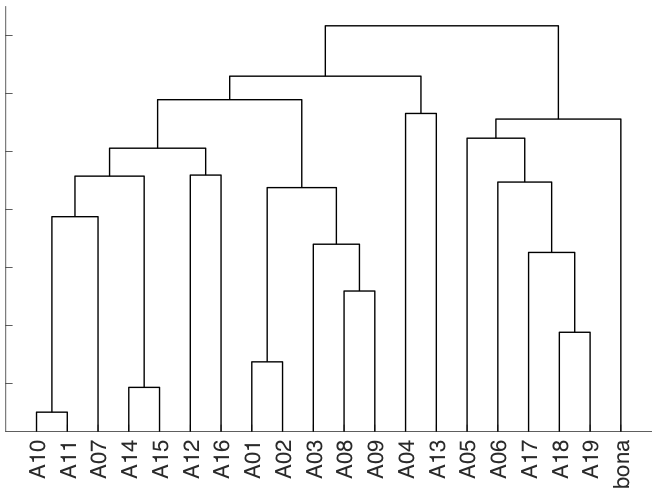}
	\caption{Agglomerative clustering of LA attacks using same input data as Figure~\ref{fig:visualization_xvector}.}
	\label{fig:LA_dendrogram}
\end{figure}

Figure~\ref{fig:visualization_xvector} is a scatter plot of the processed x-vectors. The different attacking systems and bona fide data can be identified by the colors. The circle roughly indicates the range of mean $\pm 3$ standard deviation for the bona fide data. While some of the attacking systems are well separated from the bona-fide data, spoofed data from A04, A16, A12, A11, and A10 overlapped with the bona fide data. For example, A04 and A16 are expected to be located near the bona fide speech because they are unit-selection-based TTS systems and can preserve the acoustic features of the bona-fide data in the spoofed data to some degree. A11 and A10 are also close to bona-fide, and they also overlapped with each other. This is reasonable because A11 and A10 share the same TTS architecture except for the waveform generation modules.

Besides the 2D visualization, we carried out a clustering process to find out which attacks are naturally grouped together. To this end, we used the same (pre-processed) x-vectors to compute all pairwise distances between attacks. If we let $X_i$ and $X_j$ denote the x-vector collections of attacks $i$ and $j$ ($i,j=1,\dots,20$), we define the distance of $X_i$ and $X_j$ as
    \begin{equation}
        D(X_i,X_j) = \frac{1}{|X_i|}\sum_{x \in X_i} \min_{y \in X_j} \Big(1-s(x,y)\Big),
    \end{equation}
where $s(x,y)$ is the cosine similarity of x-vectors $x$ and $y$, and  $|X_i|$ denotes the number of x-vectors in $X_i$. We then run agglomerative clustering on the resulting $20\times 20$ distance matrix by using \emph{unweighted average distance} (UPGMA) as the cluster distance. The result, visualized as a dendrogram, is displayed in Figure~\ref{fig:LA_dendrogram}. This clustering result is also reasonable. The closest pairs (A10 and A11, A14 and A15, and A01 and A02), again, indicate pairs using the same TTS architecture except for the waveform generation modules. The right-side branches (A05, A06, A17, A18, and A19) correspond to the VC systems using human speech as source data. Interestingly, the VC systems using source data produced by TTS systems (A13, A14, A15) are grouped together with TTS systems.  

In ASVspoof 2015, one attacking system similar to A04 and A16 was included in the evaluation set (i.e., S10 in ASVspoof 2015). Compared with ASVspoof 2015, however, the LA subset for ASVspoof 2019 contains more challenging spoofed data. 
{By comparing Figure~\ref{fig:visualization_xvector} with a similar figure plotted for spoofing systems in ASVspoof 2015 (i.e., Figure 2 in \cite{wu2017asvspoof}), we can see that the ASVspoof 2019 database includes more spoofing systems that overlapped with the bona fide data in the 2D feature space. In addition to the evidence from the feature visualization, we further demonstrate in Section~\ref{S:8} that spoofed data in the ASVspoof 2019 database has high perceptual naturalness and quality. Some of the spoofed data are even challenging to detect for human beings.}
Therefore, the ASVspoof 2019 database is expected to be used to examine how CMs perform facing the advanced TTS and VC spoofing systems.

\begin{figure}[!t]
	\centering
	\includegraphics[width=1.0\linewidth]{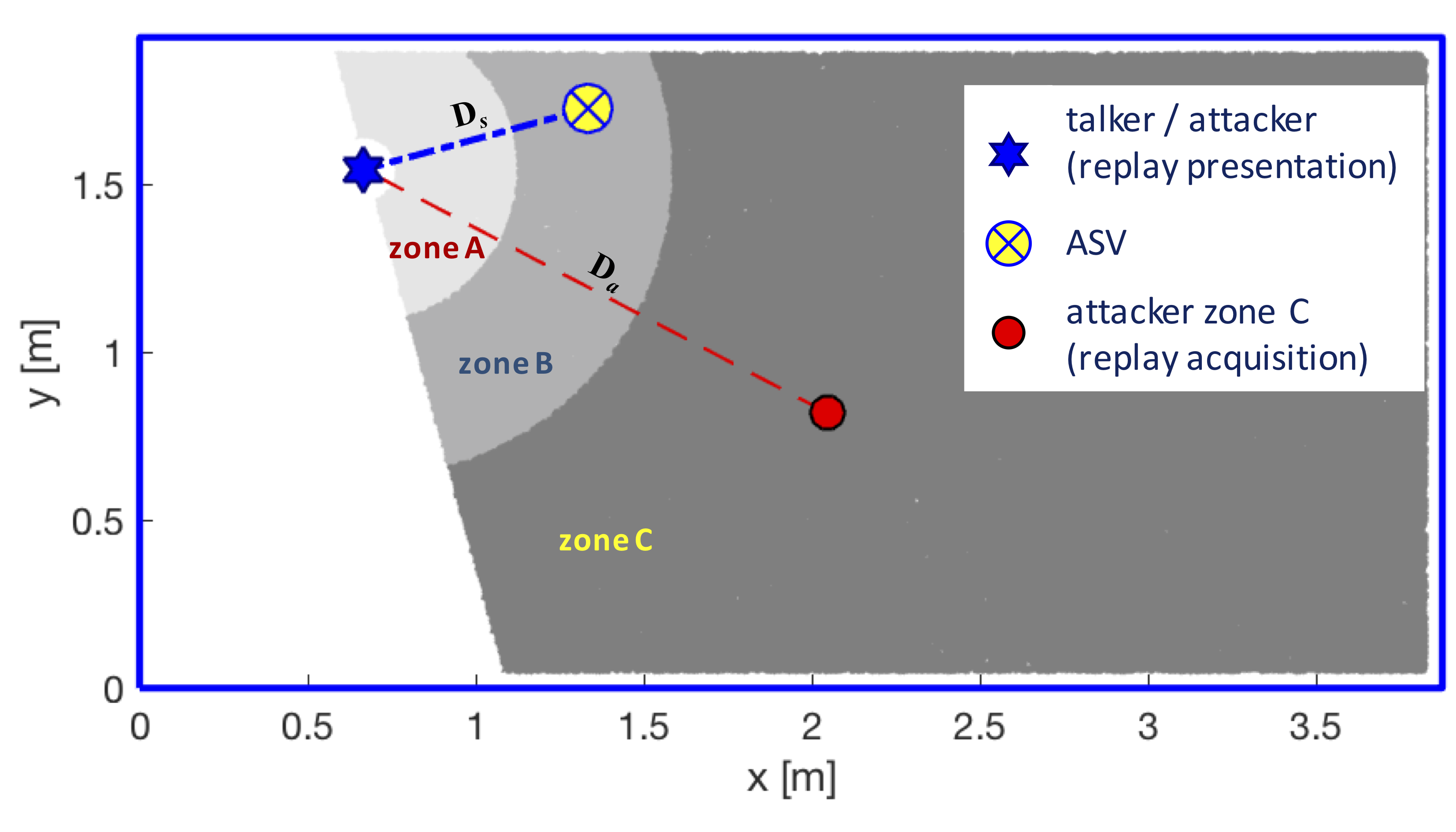}
	\caption{Illustration of ASVspoof 2019 physical access (PA) scenario.  Replay attacks are simulated within \textcolor{black}{an acoustic environment/room of dimensions $\textup{x} \times \textup{y}$} (7.5m$^2$ in the example above) with controllable reverberation.  Recordings of bona fide presentations are acquired from distance $\textup{D}_a$ from the talker.  Bona fide or replay presentations are made at distance $\textup{D}_s$ from ASV microphone.}
	\label{fig:physicalScenario}
\end{figure}

\section{Details of PA subset}
\label{S:4}

In the physical access (PA) scenario, spoofing attacks are assumed to be presented at the sensor/microphone level.  In this scenario, the microphone is a fixed component of the ASV system.
There is still variability in the channel, however.  Prior to acquisition, speech signals propagate through a physical space in which the position of the speaker and microphone can vary.

Since they are presented at the sensor level and cannot be injected post-sensor, all spoofing attacks in this scenario are referred to as replay attacks.
In contrast to the LA scenario, spoofing attacks in the PA scenario conform to the ISO definition of \emph{presentation attacks}~\cite{ISOpresentationAtack}.  
Replay attacks involve the presentation to the ASV microphone of surreptitiously captured recordings of bona fide access attempts.  
The PA scenario is thus relevant not just to ASV but also to the detection of fake audio, a problem that plagues a host of other applications such as voice interactive devices and services, e.g.,\ smart-speakers and voice-driven assistants.

%

Whereas the 2017 ASVspoof database consists of \emph{real} replay recordings, the 2019 edition consists of \emph{simulated} replay recordings \textcolor{black}{in a manner similar to~\cite{janicki2016assessment}} \footnote{{We have released another data set that was recorded and replayed in real rooms (\url{https://www.asvspoof.org/user/register}). But this real replayed data set is irrelevant to the published results of the ASVspoof challenge 2019. Neither is it included in the released ASVspoof 2019 database.}}.  Simulation was chosen in order to collect a large database of replay recordings with a carefully controlled setup.  Variability in the setup consists of differences in the acoustic environment, and the recording and presentation devices.  Simulation allows one parameter to be changed at a time while keeping everything else constant.  This approach supports a more insightful study of the different factors affecting both the impact of replay attacks on ASV\textcolor{black}{,} as well as the reliability of countermeasures.  The simulation setup is described here.

\subsection{Acoustic environment}
\label{S:4.1}

The PA scenario assumes the use of ASV within an acoustic environment such as that illustrated in Figure \ref{fig:physicalScenario}.
The acoustic configuration in which the ASV system is situated (or any other device that is tasked with determining whether a recording is bona fide or spoofed/fake) is of size~$S\textup{m}^2$.
As illustrated in Table~\ref{tab:spoofing_technique_env}, room sizes are categorized into three different intervals:
(a) small rooms of size 2-5 $\textup{m}^2$, (b)~medium rooms of size 5-10 $\textup{m}^2$, and (c) large rooms of size 10-20 $\textup{m}^2$. 
\textcolor{black}{
}

The position of the ASV/device microphone (cardioide) is illustrated by the yellow\textcolor{black}{, crossed} circle in Figure~\ref{fig:physicalScenario}.  
The position of a speaker, hereafter referred to as a talker (in order to avoid potential confusion with the loudspeaker used to mount replay spoofing attacks), is illustrated by the blue star.  
Bona fide access attempts/presentations are made by the talker when positioned at a distance $\textup{D}_s$ from the microphone.
As also illustrated in Table~\ref{tab:spoofing_technique_env}, there are three categories of talker-to-ASV distance $\textup{D}_s$: (a)~short distances of 10-50 cm, (b)~medium distances of 50-100 cm, and (c)~large distances of between 100 and 150 cm. 

Each physical space is assumed to exhibit reverberation variability according to the differences between spaces, e.g., the wall, ceiling, and floor absorption coefficients, as well as the position in the room.
The level of reverberation is specified in terms of the T60 reverberation time denoted by~$\textup{R}$.
As per Table~\ref{tab:spoofing_technique_env}, \textcolor{black}{there are three categories of T60 values: (a)~short, with a T60 time of 50-200 ms; (b)~medium, with a T60 time of 200-600 ms; (c)~high with a T60 time of 600-1000 ms.}

\begin{table}[!t]
\caption{\label{tab:spoofing_technique_env} Environment is defined as  triplet (S,R,$\textup{D}_s$), \textcolor{black}{each element of which takes one value in set (a,b,c) as a categorical value.}}
\begin{center}
\small
\tabcolsep 4.5pt
\begin{tabular}{|l|c|c|c|}
\hline
\multirow{2}{*}{ Environment definition } & \multicolumn{3}{c|}{ labels } \\
& a & b & c \\
\hline\hline
S: Room size ($\textup{m}^2$) & 2-5 & 5-10 & 10-20 \\
R: T60 (ms) & 50-200 & 200-600 & 600-1000 \\
$\textup{D}_s$: Talker-to-ASV distance (cm) & 10-50 & 50-100 & 100-150 \\
\hline
\end{tabular}
\end{center}
\end{table}

While they are not specific parameters of the setup, the position of the ASV system microphone and the talker can all vary within the physical space.
Their positions are set randomly within the room according to the category of room size $\textup{S}$ and the talker-to-ASV distance $\textup{D}_s$. 
The talker is, however, always assumed to speak in the direction of the microphone.

\subsection{Replay devices}
\label{S:4.2}

The manner by which replay attacks are mounted is also illustrated in Figure \ref{fig:physicalScenario}.  
A replay spoofing attack is mounted by (i)~making a surreptitious recording of a bona fide access attempt and (ii)~presenting the recording to the ASV microphone.  
Attackers acquire recordings of bona fide access attempts at the position indicated by the red circle in Figure~\ref{fig:physicalScenario} at distance $\textup{D}_a$ from the talker.
For a given room, their subsequent presentation to the microphone is nonetheless made from the same distance $\textup{D}_s$ as bona fide access attempts.  

Recordings are assumed to be made in one of three \emph{zones} illustrated in Figure \ref{fig:physicalScenario}, each representing a different interval of distances $\textup{D}_a$ from the talker.  
\textcolor{black}{They are illustrated in Table~\ref{tab:spoofing_technique_att}: (A)~where $\textup{D}_a$ is 10-50 cm, (B)~where $\textup{D}_a$ is 50-100 cm, and (C)~where $\textup{D}_a$ is over 100 cm.}
Recordings captured in Zone A (nearest to the talker) are expected to be of higher quality (higher signal-to-reverberation ratio) than those made in zones B and C further away from the talker.

\begin{table}[t]
\caption{\label{tab:spoofing_technique_att} \textcolor{black}{Replay attack is defined as duple ($\textup{D}_a$,Q), each element of which takes one value in set (A,B,C) as a categorical value.}}
\begin{center}
\small
\begin{tabular}{|l|c|c|c|}
\hline
\multirow{2}{*}{ Attack definition } & \multicolumn{3}{c|}{ labels } \\
& A & B & C \\
\hline\hline
$\textup{D}_a$: Attacker-to-talker distance (cm) & 10-50 & 50-100 & $>$ 100 \\
Q: Replay device quality & perfect & high & low \\
\hline
\end{tabular}
\end{center}
\end{table}

\begin{table}[t]
\caption{\label{tab:spoofing_technique_att2} \textcolor{black}{Definition of replay device quality (Q). OB refers to occupied bandwidth, minF is lower bound of the OB, LNLR is  linear-to-non-linear power ratio.}}
\begin{center}
\small
\begin{tabular}{|c|c|c|c|}
\hline
Replay device quality & OB (kHZ) & minF (Hz) & \textcolor{black}{LNLR (dB)} \\
\hline\hline
Perfect & inf & 0 & inf \\
High & $>$ 10 & $<$ 600 & $>$ 100 \\
Low & $<$ 10 & $>$ 600 & $<$ 100 \\
\hline
\end{tabular}
\end{center}
\end{table}

\begin{figure*}[!t]
	\centering
	\includegraphics[width=0.75\linewidth]{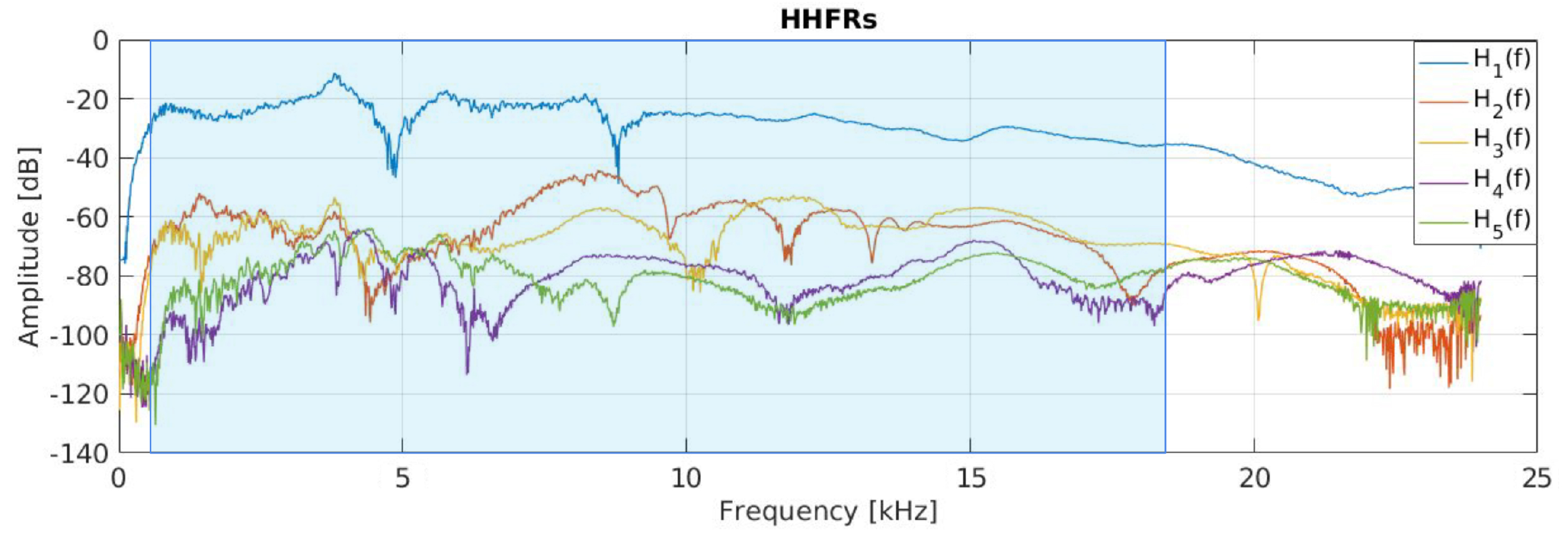}
	\vspace{-5mm}
	\caption{
	Illustration of set of higher harmonic frequency responses (HHFRs) for arbitrary smart-tablet device estimated using  synchronized-swept-sine approach to nonlinear system identification based on \emph{nonlinear convolution}. 
	$H_1$ is linear component, while $H_2 - H_5$ are higher order non-linear components. 
	Blue/shaded region is occupied bandwidth, difference in frequency between points where integrated power crosses 0.5\% and 99.5\% of total power in spectrum.
	}
	\label{fig:deviceIRs}
\end{figure*}

In addition to reverberation, replay attacks will reflect artifacts stemming from recording and presentation devices.  
Recordings are made with a microphone \textcolor{black}{which is different to that of the ASV system.}  
Presentations are made using a loudspeaker.  
The behavior of even miniature microphones can be mostly linear, and their frequency response can be relatively flat.
Loudspeakers, in contrast, often exhibit non-linear behavior and rarely have flat frequency responses.
{In this case, and as is usual in situations involving artifacts from both microphones and loudspeakers, e.g., acoustic echo cancellation, loudspeaker-related artefacts are assumed to dominate~\cite{toole2008sound}, so microphone artifacts can be safely ignored.}
Typical non-linear and band-limiting loudspeaker effects are therefore also simulated.

Loudspeaker artifacts are included according to a generalized polynomial Hammerstein model (GPHM), the parameters of which are estimated using the Synchronized Swept Sine tool\footnote{\url{https://ant-novak.com/pages/sss/}}.
Using a GPHM, both the linear and non-linear characteristics of a loudspeaker can be modelled and then simulated using \emph{linear} impulse responses.
These are referred to as higher \textcolor{black}{harmonic} frequency responses (HHRFs).
An example set of HHRFs for an arbitrary smart-tablet device is illustrated in Figure~\ref{fig:deviceIRs}.
It shows the dominant, 1st order, linear impulse response $H_1(f)$ towards the top with those of the less significant 2nd to 5th order non-linear components $H_2(f)$...$H_5(f)$ towards the bottom.
In practice, one need only model the first five components in order to capture the most significant device non-linearities.
Also illustrated in Figure~\ref{fig:deviceIRs} by the blue/shaded box is the occupied bandwidth of the device, which extends from approximately 0.5-18kHz.
The non-linear behavior of a loudspeaker can then be simulated by convolving a replay recording with the full set of impulse responses. 

As illustrated in Table~\ref{tab:spoofing_technique_att2},
the simulations consider three categories of loudspeaker, each with a different range of occupied bandwidth (OB), minimum frequency (minF), and linear-to-non-linear power ratio (LNLR).
The first, hypothetical $\emph{perfect}$ category represents an entirely linear, full bandwidth loudspeaker.
High-quality loudspeakers have an occupied bandwidth exceeding 10 kHz and an LNLR exceeding 100 dB. 
Low-quality loudspeakers have an occupied bandwidth below 10 kHz and an LNLR below 100 dB.
Differences in bandwidth and linearity are summarized according to a single device quality indicator Q. 
The correspondence between replay device quality and Q indicator is illustrated in the last row of Table~\ref{tab:spoofing_technique_att}. 
Replay attacks mounted with low-quality loudspeakers are expected to be detected with relative ease, whereas those mounted with the `\textcolor{black}{perfect}' loudspeaker represent the worst case scenario.

\begin{table}[t!]
\renewcommand{\arraystretch}{1}
\caption{\textcolor{black}{List of real devices from which measurements were taken for simulation of replay attack presentation. Q indicates device quality (B high, C low).  Device code signifies device type: bluetooth (BT); headphone (H); mobile smartphone (M);  larger consumer and professional loudspeaker (LS); tablet (T); laptop (LT). Level indicates volume (high, low) used during device characterisation.  Right-most column indicates whether measured device characteristics were used for the simulation of utterances in training and development set (known devices) or evalutaion set (unknown devices).} 
}
\centering
\scalebox{0.8}{
\begin{small}
\begin{tabular}{|c|c|c|c|c|}
\hline
Q & device & model and brand & level & known \\
\hline
\hline
\multirow{10}{*}{ B } & BT2 & EC technology S10 & low & x \\
 & BT4 & Sony SRS XB3 & low & x \\
 & H1 & Beyerdynamic DT770 PRO & low & x \\
 & H2 & Beyerdynamic DT770 PRO & high & x \\
 & M10 & iPhoneSE & low & x \\
 & LS4 & Desktop speaker & low & x \\
 & LS1 & Behringer Truth B2030A & high &  \\
 & LS2 & Behringer Truth B2030A & low &  \\
 & LS5 & ESI nEar08 Ex & high &  \\
 & LS6 & ESI nEar08 Ex & low &  \\
\hline
\multirow{30}{*}{ C } & BT1 & EC technology S10 & high & x \\
 & BT3 & Sony SRS XB3 & high & x \\
 & LS3 & Desktop speaker & high & x \\
 & LT3 & DELL Vostro V131 & high & x \\
 & LT4 & DELL Vostro V131 & low & x \\
 & M11 & Motorola Mot G6 plus & high & x \\
 & M12 & Motorola Mot G6 plus & low & x \\
 & M15 & Oneplus 5T & high & x \\
 & M16 & Oneplus 5T & low & x \\
 & M19 & Oppo F7 & high & x \\
 & M1 & BQ AQUARIS E5 & high & x \\
 & M20 & Oppo F7 & low & x \\
 & M2 & BQ AQUARIS E5 & low & x \\
 & M3 & Huawei P10 lite & high & x \\
 & M4 & Huawei P10 lite & low & x \\
 & M5 & Huawei P10 plus & high & x \\
 & M9 & iPhoneSE & high & x \\
 & T1 & Samsung Galaxy TabA & high & x \\
 & T2 & Samsung Galaxy TabA & low & x \\
 & BT5 & UE BOOM2 & high &  \\
 & BT6 & UE BOOM2 & low &  \\
 & LT1 & Alienware & high &  \\
 & LT2 & Alienware & low &  \\
 & M13 & Homtom HT26 & high &  \\
 & M17 & Oneplus One & high &  \\
 & M18 & Oneplus One & low &  \\
 & M22 & Xiaomi MI5 & high &  \\
 & M23 & Xiaomi MI5 & low &  \\
 & M25 & Xiaomi redmi note3 & low &  \\
 & M8 & iPhone4 & low &  \\
\hline
\end{tabular}
\end{small}
}
\label{Table:PA_Reply_device}
\end{table}

A list of real loudspeaker devices is shown in Table~\ref{Table:PA_Reply_device}.  
{The devices with a Q indicator equal to B and C correspond to categories High and Low in Table~\ref{tab:spoofing_technique_att2}, respectively.}
They include a variety of smaller bluetooth (BT), headphones (H), mobile (M), tablet (T) and laptop (LT) loudspeakers, in addition to larger consumer and professional loudspeakers (LS).  Each device was set to operate at either an arbitrary low or high volume level.  Due to the mechanical operation of typical loudspeakers, \emph{e.g.,}~the movement of the voice coil towards the pole piece, higher volume levels typically lead to more significant non-linearities. This translates into higher energy for higher order harmonics, \emph{e.g.,} $H_2(f)$-$H_5(f)$.  The right-most column on Table~\ref{Table:PA_Reply_device} shows whether the devices were used to simulate replay utterances in the training and development partitions, in which case they are \emph{known} devices, or whether they were used for the simulation of replay utterances in the evaluation set, in which case they are \emph{unknown} devices.

\begin{figure*}[!t]
	\centering
	\includegraphics[width=0.9\linewidth]{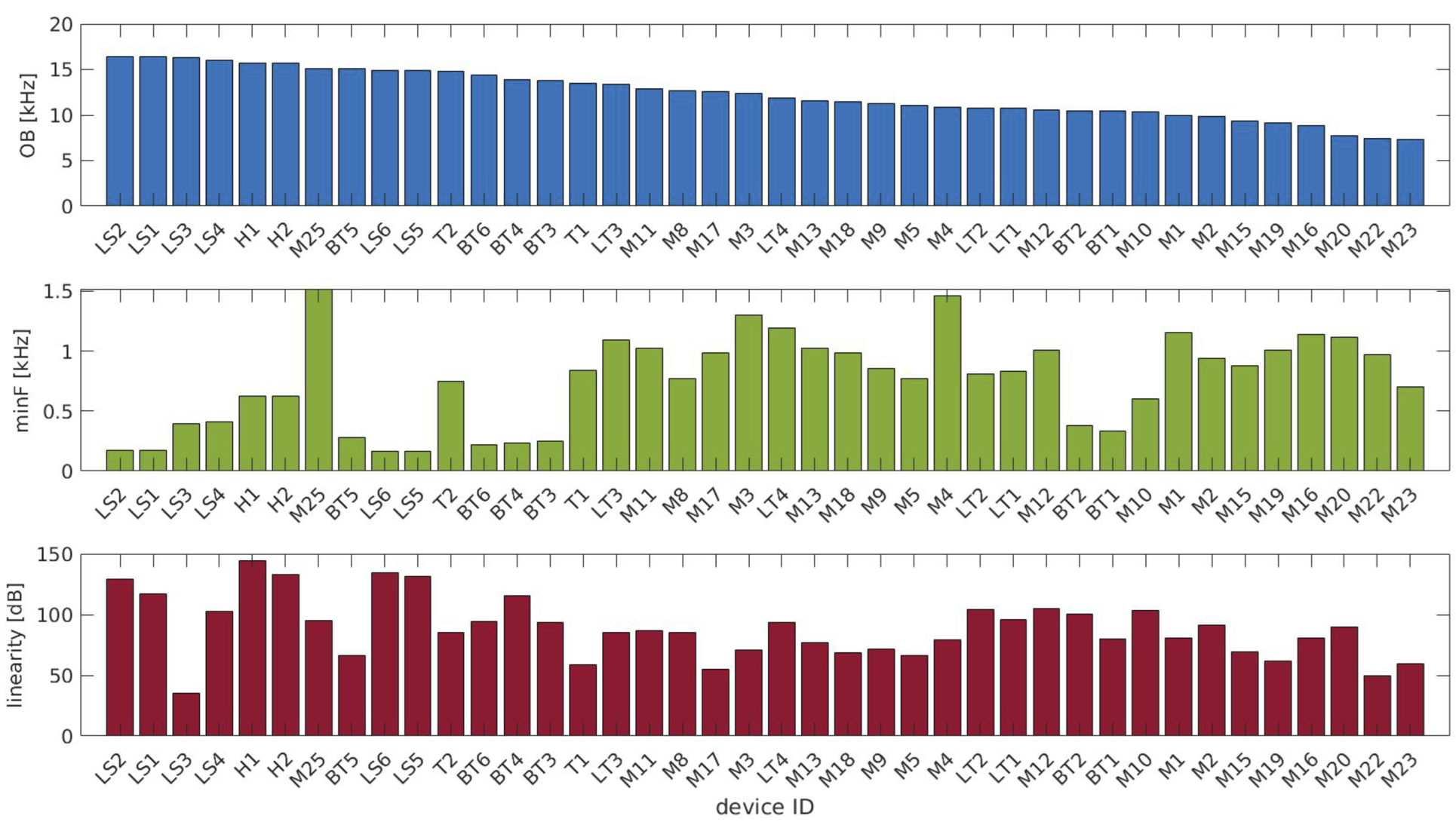}
	\caption{Characteristics measured from 40 different loudspeaker devices listed in Table~\ref{Table:PA_Reply_device}.  The top plot shows the operational bandwidth (OB).  The middle plot shows the lower bound of the OB (minF).  The bottom plot shows the linear-to-non-linear power ratio (LNLR) in the range of the OB.
	\textcolor{black}{Change label of lower plot to LNLR.}}
	\label{fig:deviceQ}
\end{figure*}

\textcolor{black}{An illustration of the occupied bandwidth (OB), minimum frequency (minF) and linear-to-non-linear power ratio (LNLR) values for each device is presented in Figure~\ref{fig:deviceQ}.  Devices are sorted according to the OB for which values range between approximately 7.5 and 17kHz.  The minF ranges from approximately 0.2kHz to 1.5kHz, whereas the LNLR ranges from 30 to 145dB.  The set of devices represent a broad range of device types, characteristics and qualities that could be used for the mounting of replay devices.}

\subsection{Simulation procedure}
\label{S:4.3}


The approach used to simulate room acoustics is that described in~\cite{Allen79}.
Simulations were done using Roomsimove\footnote{\url{http://homepages.loria.fr/evincent/software/Roomsimove_1.4.zip}}~\cite{vincent08}, which takes into account the entire acoustic environment, including room size, reverberation, and varying source/receiver positions, which includes source directivity.
The same software and technique has been applied successfully for data augmentation in well-known x-vector-based speaker and speech recognition recipes~\cite{Snyder2018XVectorsRD,Ko2017ASO}.

\begin{figure}[!t]
	\centering
	\includegraphics[width=1.0\linewidth]{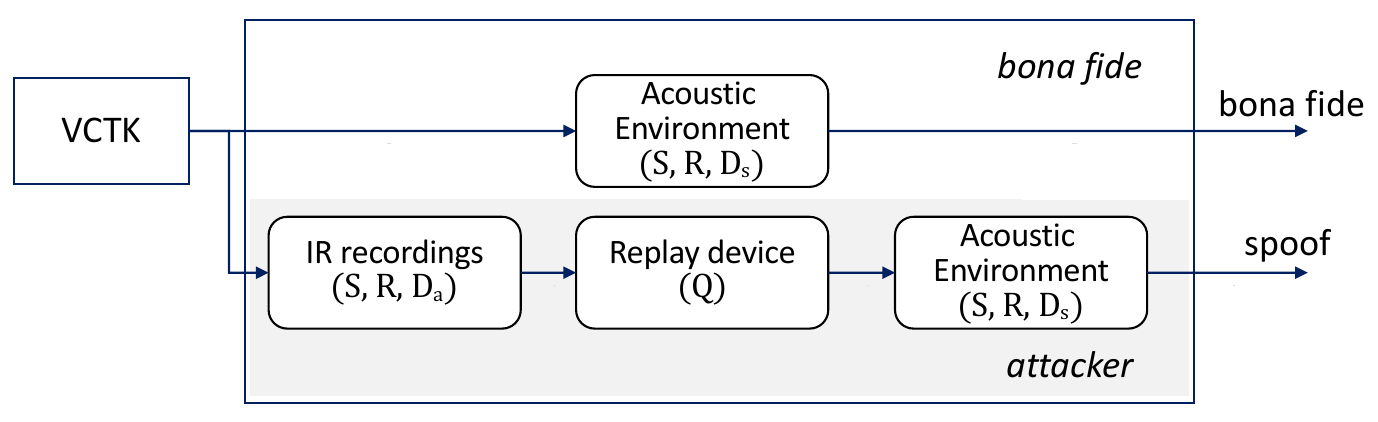}
	\caption{Illustration of PA simulation process based on impulse response (IR) modeling approach.
	Simulations take into account size of acoustic environment S and level of reverberation R. 
	Bona fide access attempts are made at distance $\textup{D}_s$ from ASV microphone, whereas surreptitious recordings are made at distance $\textup{D}_a$ from talker before being presented to ASV microphone, also at distance $\textup{D}_s$. \textcolor{black}{Effects of digital-to-analogue conversion, signal amplification and replay (using loudspeaker) are all modelled,  
	and represented with single device quality indicator Q.} 
	}
	\label{fig:bf_rep_conf}
\end{figure}




As illustrated in the bottom panel of Figure~\ref{fig:database_protocol}, bona fide and replay utterances were simulated\footnote{Note that, in the defined scenario, both bonafide and spoofed access attempts occur in the same environment (the one in which the ASV system is placed). For this reason, to generate bonafide speech, the source clean speech (from the VCTK corpus, recorded in an anechoic room) is also processed with environment simulator.}
in 27 different environments.  
An environment was defined according to a combination of room size S, reverberation level R, and talker-to-ASV distance $\textup{D}_s$.  
With each parameter being categorized into three intervals as per Table~\ref{tab:spoofing_technique_env}, a single acoustic environment is specified according to a tuple (SR$\textup{D}_a$).  
By way of example, environment `aaa' corresponds to a small room size of 2-5 $\textup{m}^2$ with a T60 reverberation of 500-200 ms and a talker-to-ASV distance of 10-50 cm.
The three parameters, each categorized into three intervals, gives the full set of 27 different acoustic environments: aaa, aab, aac,... ccb, ccc.

In each acoustic environment, replay utterances are simulated according to nine different attack types.
They depend upon the attacker-to-talker distance {$\textup{D}_a$} and the replay device quality~Q.
With each parameter being categorized into the three intervals illustrated in Table~\ref{tab:spoofing_technique_att}, each attack is specified by the {duple ($\textup{D}_a$Q)} so that attack type `AA' corresponds to an attacker-to-talker distance of 10-50 cm and presentations made with the hypothetical \emph{perfect} loudspeaker.
The two different parameters, each categorized into three different intervals, give the nine different replay configurations: AA, AB, AC,... CB, CC.

The full simulation procedure is illustrated in Figure~\ref{fig:bf_rep_conf}.  
By combining the 9 different replay configurations with the 27 different acoustic configurations, there was a total of 243 different evaluation conditions.
For maximum precision, all simulations were performed with \textcolor{black}{the original VCTK data whose sampling rate is 96 kHz}. 
For consistency with the LA setup, the resulting 96-kHz data was then downsampled to 16kHz with 16 bits-per-sample resolution.
Bona fide presentations were generated from the simulation of acoustic environment effects only (SR$\textup{D}_s$).
In contrast, replay attacks were generated from the simulation of recording effects (SR$\textup{D}_a$), loudspeaker effects (Q), and replay presentation (SR$\textup{D}_s$).
All simulations were made with the Roomsimove default microphone and talker heights of 1.1m. The height of each room is set to the Roomsimove default of 2.7 m.

{The quality of the loudspeakers (Q) has been defined as the best combination of OB, minF and LNLR. The measurements of the 40 devices/loudspeakers has been performed in a quasi-anechoic room of 2 $\textup{m}^2$. The distance between the loudspeaker and the microphone is set to 1 m. For all the measurements, we have used the AKG C3000 flat-response condenser microphone and the Focusrite Scarlet 2i2 low-noise audio device. For each loudspeaker, we set up two different levels: low and high, meaning that the volume of the device is set to 1/4 and 3/4, respectively}.






\subsection{Visualization of PA subset}

\begin{figure}[!t]
	\centering
	\includegraphics[width=1.0\linewidth]{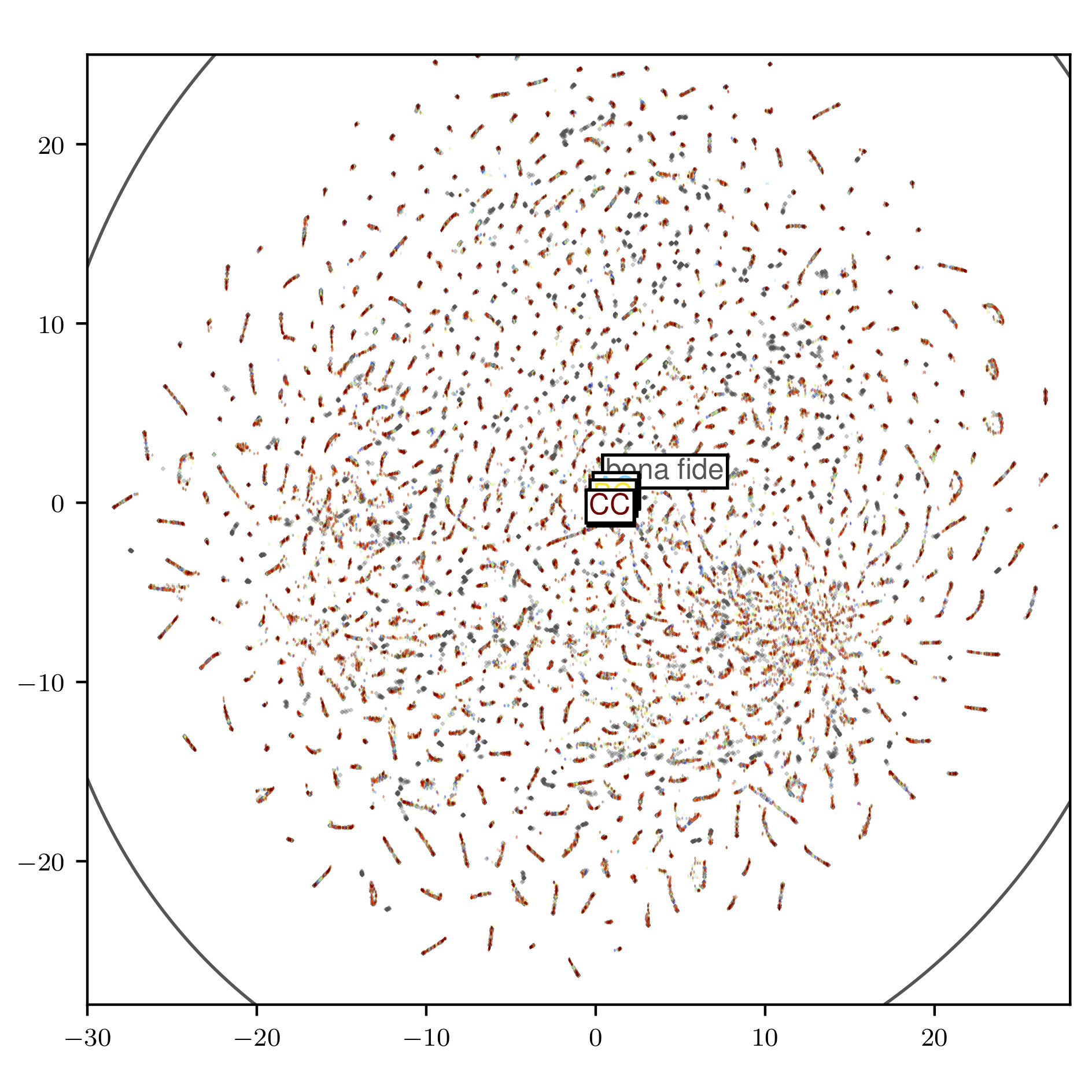}
	\caption{
	As for Figure~\ref{fig:visualization_xvector}, except for PA data.  Bona fide utterances and replayed versions according to 9 different replay configurations are completely overlapping.  Mini-clusters correspond to speaker-utterance combinations.
	}
	\label{fig:visualization_xvector_PA}
\end{figure}

\textcolor{black}{Figure~\ref{fig:visualization_xvector_PA} shows a visualisation of the PA data in the same way as Figure~\ref{fig:visualization_xvector} showed for LA data.  The plots are very different, however.  Each colour in Figure~\ref{fig:visualization_xvector_PA} corresponds to replay utterances in each replay configuration, \emph{e.g.,} AA...CC, in addition to bona fide utterances.  Each mini-cluster corresponds to a different speaker-utterance combination.}

\textcolor{black}{In contrast to the same plot for the LA scenario, the impact of replay attacks is not at all evident in Figure~\ref{fig:visualization_xvector_PA}; the distribution of replay and bona fide utterances is completely overlapping.  This is hardly surprising, however, since x-vector representations attenuate channel variability.  Since channel artefacts are essentially the only means to distinguish between bona fide and replay utterances, the x-vector representation of each are then highly alike.  This observation shows the potential difficulty in detecting replay attacks and also the need to develop \emph{independent} ASV and CM systems.  It may also suggest that the latter should use employ an utterances representation that emphasizes, rather than attenuates channel variability.} 


\section{ASV and CM baseline systems}
\label{S:5}


As for the 2017 edition, countermeasure (CM) baseline systems were provided to all ASVspoof 2019 participants.  
In addition, participants were also provided with ASV scores for all bona fide and spoofed trials.  
For the challenge itself, these were provided for the training and development partitions only.  
The now-public version of the database also contains ASV scores for the evaluation partition.  
ASV scores, produced by a fixed ASV system designed by the organizers, support the use of the t-DCF metric \cite{Kinnunen2018-tDCF}, which assesses the impacts of both spoofing and CMs on ASV performance.  
Participants may then focus exclusively on the design of CMs without the need to develop and optimize an additional ASV system.  
The baseline ASV and CM systems are described here.

\subsection{ASV baseline}
\label{Section:x_vector}

The ASV system uses DNN-based \emph{x-vector} speaker embeddings \cite{Snyder2018XVectorsRD} together with a \emph{probabilistic linear discriminant analysis} (PLDA) \cite{prince2007probabilistic} backend. 
The x-vector extractor is a pre-trained\footnote{\url{http://kaldi-asr.org/models/m7}} neural network available for the Kaldi~\cite{povey2011kaldi} toolkit. 
The network is trained with MFCC features extracted from audio data from 7,325 speakers of the VoxCeleb1 and VoxCeleb2~\cite{nagrani2017voxceleb} databases. 
The x-vector model consists of a 5-layer deep \emph{time-delay neural network} (TDNN), followed by statistics pooling and two fully connected layers before a softmax output. The statistics pooling layer converts the frame-level TDNN output to utterance-level representations by computing the mean and standard deviation of features over time. 
The x-vector embeddings are obtained from the first fully connected layer after the pooling layer. 
The dimension of the embedding layer is 512, and the embeddings are extracted without the application of the ReLU activation function or batch normalization of the embedding layer. 
The network was trained with a stochastic gradient descent algorithm. 
Further details concerning network parameters and data preparation are available in~\cite{Snyder2018XVectorsRD}. 

The original Kaldi recipe was modified to include PLDA adaptation using disjoint, bona fide, in-domain data. 
Bona fide CM training data (\#5 and \#17 in Figure \ref{fig:database_protocol}) was used as in-domain data for both the LA and PA scenarios. Note that adaptation was performed separately for the LA and PA scenarios since bona fide recordings for the latter only contain simulated acoustic and recording effects. 
The within-class and between-class covariances were adapted using Kaldi's domain adaptation technique (Algorithm 1 of \cite{bousquet2019robustness}) with scaling factors of $\alpha_w = 0.25$ and $\alpha_b = 0$ for LA and $\alpha_w = 0.90$ and $\alpha_b = 0$ for PA.
The scaling factors were optimized to deliver the best recognition performance with non-target, zero-effort imposters.

The x-vector representations of the enrollment utterances were averaged to create a single x-vector per enrolled speaker. 
Before PLDA-based log-likelihood-ratio scoring, x-vectors were centered, reduced to a dimension of 200 with a linear discriminant analysis (LDA) transform that whitens the within-class covariance matrix, and normalized to the unit length. 
The Kaldi implementation of PLDA~\cite{ioffe2006probabilistic} is used for scoring.

\subsection{CM baseline}
Two baseline CM systems were made available to ASVspoof 2019 participants. 
Both use a common Gaussian mixture model (GMM) back-end classifier with either \emph{constant-Q cepstral coefficient} (CQCC) features~\cite{CQCC2016,TODISCO2017516}~(B01) or \emph{linear frequency cepstral coefficient} (LFCC) features~\cite{Sahidullah15}~(B02).

CQCC baseline B01 uses a constant-Q transform (CQT), which is applied with a maximum frequency of $f_{nyq}=f_s/2$, where $f_s=16 \textup{kHz}$ is the sampling frequency. 
The minimum frequency is set to nine octaves below the maximum frequency $f_{min}=f_{max}/2^9\simeq15 \textup{Hz} $.
The number of bins per octave is set to~96. 
The resulting geometrically-scaled CQT spectrogram is re-sampled to a linear scale using a sampling period of~16. 
The discrete cosine transform (DCT) is then applied to obtain a set of static cepstral coefficients. 
The full set of CQCC features includes 29+0th order static coefficients plus corresponding delta and delta-delta coefficients computed using two adjacent frames.

LFCC baseline B02 uses a short-term Fourier transform.  
The input signal is the first frame blocked with a 20~ms Hamming window and a 10~ms shift. 
The power magnitude spectrum of each frame is calculated using a 512-point FFT. 
A triangular, linearly spaced filterbank of 20 channels is then applied to obtain a set of 20 coefficients. 
LFCC features are obtained by applying the DCT to the filterbank log-energy densities.
The set of 19+0th static coefficients are then augmented with corresponding delta and delta-delta coefficients, again computed using two adjacent frames.

The back-end for both B01 and B02 use a pair of GMMs each with 512 components. 
Parameters for bona fide and spoofed speech models are trained separately with 20 iterations of the expectation-maximization (EM) algorithm. 
Finally, scores are the log-likelihood ratio between the two hypotheses, namely that a given trial is either bona fide or spoofed speech. 
Baseline CMs are trained separately for LA and PA scenarios using designated CM training data.

\section{ASV and CM baseline results}
\label{S:6}
\textcolor{black}{This section describes the results obtained with the ASVspoof 2019 ASV and CM baselines. The results are presented separately for the LA and PA datasets. The metrics are the equal error rate (EER) and new ASV-centric tandem detection cost function (t-DCF)~\cite{Kinnunen2018-tDCF} which combines the errors of both ASV and CM systems into a single metric.} 

\subsection{Analysis of LA subset}

\subsubsection{Analysis of baseline ASV performance}

First, we assessed the impact of the different attacks with the baseline ASV system. In Table \ref{Table:LA_ASV_Baseline}, we show the ASV performance for the  LA subset using the x-vector system described in Section~\ref{Section:x_vector}. The performance was measured in terms of EER computed with a tool developed for the challenge\footnote{\url{https://www.asvspoof.org/asvspoof2019/tDCF_matlab_v1.zip} (MATLAB) \url{https://www.asvspoof.org/asvspoof2019/tDCF_python_v1.zip} (Python)}. The first row of the results shows the performance with the bona fide imposters, whereas the rest of the rows show the results for all 19 attacks, i.e., A01-A19, split into development and evaluation sets. 

\begin{table}[t!]
\renewcommand{\arraystretch}{1}
\caption{ASV performance in terms of EER (\%) on LA subset of ASVSpoof 2019 dataset for baseline and different attack conditions for development and evaluation set. Note that A16 used same TTS algorithm as A04, and A19 used same VC algorithm as A06.}
\centering
\begin{small}
\begin{tabular}{|c|c|c|}
\hline
Attack & Development & Evaluation   \\
\hline
Baseline      &    2.43	& 2.48        \\
\hline
A01    &   24.52    &    -    \\
A02    &   15.04    &    -    \\
A03    &   56.94    &    -    \\
A04    &   63.02    &   -     \\
A05    &  21.90     &   -     \\
A06    & 10.11      &    -    \\
A07    &   -    &    59.68    \\
A08    &   -    &    40.39    \\
A09    &   -    &   8.38     \\
A10    &  -     &   57.73     \\
A11    &  -     &   59.64     \\
A12    &   -    &   46.18     \\
A13    &   -    &   46.78     \\
A14    &  -     &   64.01     \\
A15    &   -    &   58.85     \\
A16    &   -    &   64.52     \\
A17    &  -     &   3.92     \\
A18    &   -    &  7.35      \\
A19    &  -     &  14.58      \\
\hline
\end{tabular}
\end{small}
\label{Table:LA_ASV_Baseline}
\end{table}

The results indicate that the spoofed data created with various VC and TTS methods drastically degraded the ASV performance. However, the severity of the attacks varied across different methods. For instance, A04, the TTS attack created with the waveform concatenation method, increased the EER from 2.43\% to 63.02\%, and it turned out to be the most severe attack in the development set. Apart from this, A03, another attack in the development set created with the feedforward-neural-network-based acoustic model and the WORLD vocoder also severely degraded the ASV performance. Both A03 and A04 yielded an EER of more than 50\%, which also indicates that the synthetic speech created with these two attacks was more similar to the speakers than the genuine target sentences used, interestingly. Note that all the spoofing methods used 200 utterances per target speaker to train the spoofing system (as illustrated in Figure \ref{fig:database_protocol}), and these two methods seem to be more efficient in cloning speaker-related characteristics closer to the enrollment utterances. In comparison, the attacks developed with the acoustic model based on GMM-UBM (i.e., A06) displayed the least degradation in ASV performance. The other two attacks created with auto-regressive LSTM (A01 and A02) or with VAE (i.e., A05) were also able to severely degrade the ASV performance but not like A03 or A04. Nevertheless, the poorest attack in the development set (i.e., A06) also increased the baseline EER by more than four times.

\begin{figure}[!t]
	\centering
	\includegraphics[width=0.47\textwidth]{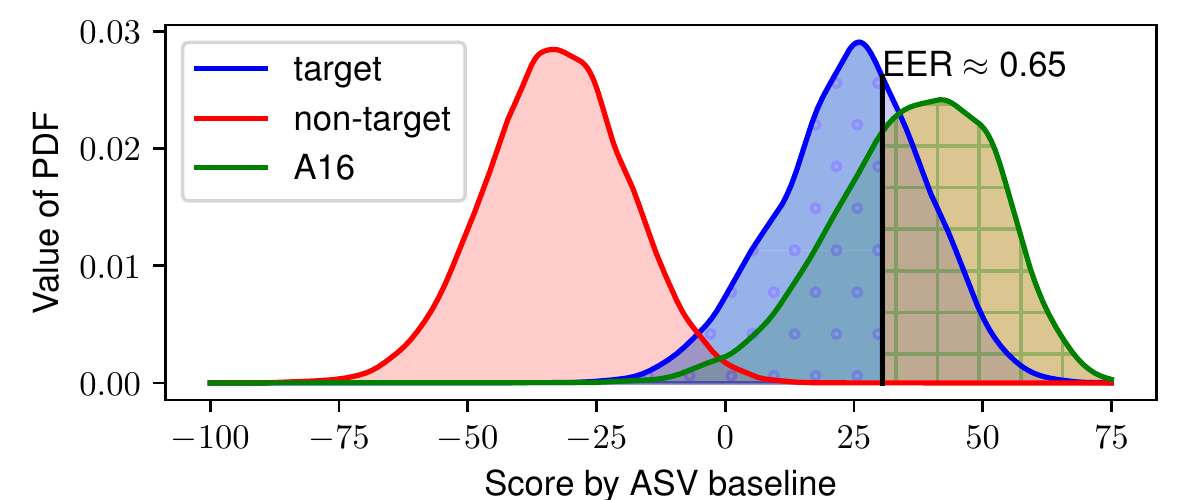}
	\caption{{ASV score distributions for target bona fide, non-target bona fide, and spoofed data from A16. Vertical black line denote classification threshold between spoofed and target bona fide data. Dot and square shaded areas correspond to false reject errors and false accept errors, respectively.}}
	\label{fig:fig_score_a16}
\end{figure}

{Note that some spoofing system such as A16 led to an ASV EER larger than 50\%. This is because the spoof data from those spoofing systems acquired ASV scores even higher than that of bona fide data, as Figure~\ref{fig:fig_score_a16} plots for the case of A16. Accordingly, the false reject rate, the false accept rate, and the EER become larger than 50\%.}

\begin{figure}[!t]
	\centering
	\includegraphics[clip, trim=10.3cm 0.2cm 1.5cm 0cm, width=1.0\linewidth]{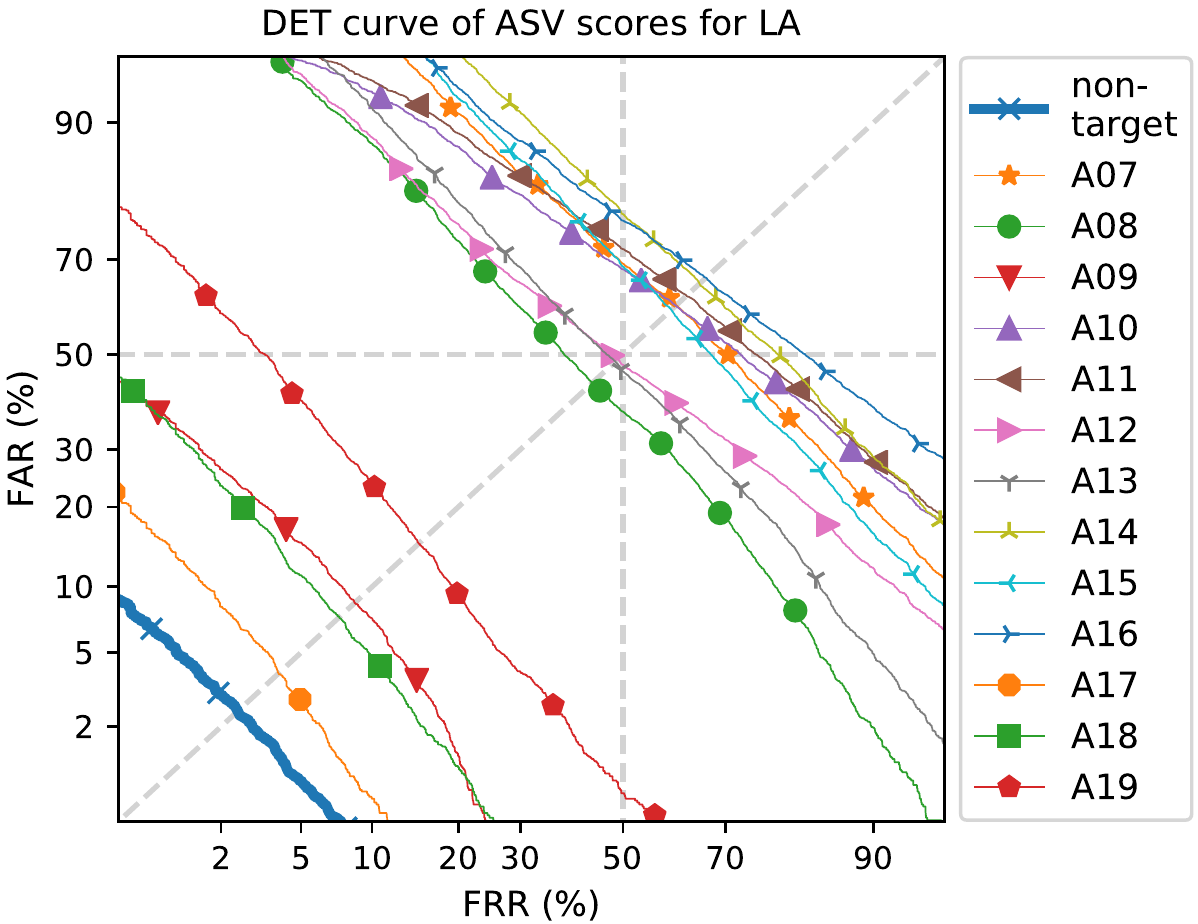}
	\caption{ASV DET curves for eval side of LA subset. DET curves are computed by assigning bonafide target scores to positive class while assigning bonafide non-target scores or scores from different attack types to negative class. These DET curves can be compared with ones in upcoming Figure \ref{fig:human_similarity} to see how ASV results compare with human perception.}
	\label{fig:la_asv_det}
\end{figure}

From the results of the evaluation set, which we depict as DET curves in Figure \ref{fig:la_asv_det}, we also observed that different attacks had different degrees of severity. Here also, the waveform concatenation based attack (i.e., A16, which is the same as A04 used in the development set) turned out to be the most critical attack with an ASV EER of 64.52\%. The spoofing method A14, which uses LSTM with the STRAIGHT vocoder, also caused a comparable degradation in EER. The other two LSTM-acoustic-model-based spoofing methods integrated with the neural waveform model (i.e., A15) or WORLD vocoder (i.e., A07) had an ASV EER of more than 50\%. However, the LSTM-based acoustic model that uses the Vocaine vocoder (i.e., A09) for waveform generation did not degrade the ASV performance like the other LSTM-based models. Amongst the other attacks, the sequence-to-sequence based methods (i.e., A10 and A11) also exhibited  an ASV EER of more than 50\%. Similar to the development set, the voice conversion systems, where the source speakers are human (i.e., A17-19), showed less degradation in ASV performance compared than other attacks. The spoofing method A17 based on the VAE-based acoustic model and waveform filtering for wave generation had an EER of 3.92\%, which is comparable to the ASV performance achieved with the bona fide imposter.

\subsubsection{Analysis of baseline CM performance}

The results so far discussed do not consider any countermeasure. Therefore, they do not reflect the actual impact of the attacks in a scenario in which the ASV system is integrated with countermeasures. Table~\ref{Table:LA_CM_Baseline_Dev} shows the performance of the joint CM and ASV system for the development set in terms of min-tDCF as well the performance of the standalone CM system in terms of EER.

We observed that the B1 or CQCC-based method had lower costs as well as error rates compared with the B2 or LFCC-based method in most cases. However, for two attacks, A03 and A05, the LFCC-based system outperformed the other. For the B1 system, A05 turned out to be the most difficult attack in terms of both performance evaluation metrics. For the B2 system, A06 was the most difficult attack. Interestingly, compared with A03 and A04, the baseline ASV EERs (Table~\ref{Table:LA_ASV_Baseline}) were not that high for A05 and A06. Furthermore, attack A04, which yielded the highest ASV EER, was accurately detected by the B1 system, though it turned out to be the second most difficult attack from the B2 system perspective. Note that the B1 or CQCC-GMM system has already demonstrated its superiority over other features, including LFCC, specially for  detecting waveform concatenation based TTS attack~\cite{TODISCO2017516}. In the present scenario, the CM system was also trained with spoofed audio data generated with the same attack, here A04. This further helped the system to more efficiently detect such attacks. As a result, though A04 seemed to be the most difficult attack from the ASV viewpoint, it was easily detected by the B1 system.


\begin{table}[t!]
\renewcommand{\arraystretch}{1.2}
\caption{Performance of integrated system in terms of min-tDCF and of standalone countermeasures in terms of EER (\%) on development set (LA subset) of ASVSpoof 2019 dataset. Results are shown for two baselines, B1 (CQCC-GMM) and B2 (LFCC-GMM), separately, combined with fixed ASV system based on x-vector. Last row describes results for ``pooled condition'' when trials from all the attacks are considered for evaluation.}
\vspace{0.1cm}
\centering
\begin{small}
\begin{tabular}{|c|c|c|c|c|}
\hline
\multirow{2}{*}{Attack} & \multicolumn{2}{|c|}{B1} & \multicolumn{2}{|c|}{B2}   \\
\cline{2-5}
       & min-tDCF & 	EER& 	min-tDCF&	EER\\
\hline
A01    &	0.0000&	0.00&     0.0005&	0.03     \\
A02    &	0.0000&	0.00&      0.0000&	0.00       \\
A03    &	0.0020&	0.08&      0.0000&	0.00        \\
A04    &	0.0000&	0.00 &      0.1016&	4.90       \\
A05    &	0.0261&	0.94&       0.0033&	0.16        \\
A06    &	0.0011&	0.03&      0.2088	&5.27        \\
\hline
Pooled &	0.0123&	0.43& 0.0663	&2.71\\
\hline
\end{tabular}
\end{small}
\label{Table:LA_CM_Baseline_Dev}
\end{table}

\begin{table}[t!]
\renewcommand{\arraystretch}{1.2}
\caption{Same as Table~\ref{Table:LA_CM_Baseline_Dev} but for evaluation set}
\vspace{0.1cm}
\centering
\begin{small}
\begin{tabular}{|c|c|c|c|c|}
\hline
\multirow{2}{*}{Attack} & \multicolumn{2}{|c|}{B1} & \multicolumn{2}{|c|}{B2}   \\
\cline{2-5}
       & min-tDCF & 	EER& 	min-tDCF&	EER\\
\hline
A07&	0.0000&	0.00& 0.3263&	12.86\\
A08&	0.0007&	0.04& 0.0086&	0.37\\
A09&	0.0060&	0.14& 0.0000&	0.00\\
A10&	0.4149&	15.16& 0.5089&	18.97\\
A11&	0.0020&	0.08& 0.0027&	0.12\\
A12&	0.1160&	4.74& 0.1197&	4.92\\
A13&	0.6729&	26.15& 0.2519&	9.57 \\
A14&	0.2629&	10.85& 0.0314&	1.22 \\
A15&	0.0344&	1.26& 0.0607&	2.22\\
A16&	0.0000&	0.00& 0.1419&	6.31 \\
A17&	0.9820&	19.62&  0.4050&	7.71\\
A18&	0.2818&	3.81&  0.2387&	3.58\\
A19&	0.0014&	0.04& 0.4635&	13.94\\
\hline
Pooled &0.2366	& 9.57 &0.2116&	8.09	\\
\hline
\end{tabular}
\end{small}
\label{Table:LA_CM_Baseline_Eval}
\end{table}

Table~\ref{Table:LA_CM_Baseline_Eval} shows the result for the evaluation set. Here, we observed that the overall performance with the pooled attacks was substantially poorer than the performance on the development set. This was due to the presence of new attacks that were not included during CM training. For instance, we observed that attacks A16 and A19, which were already included in the training set, had a lower EER and min-tDCF for the B1 system. However, for these two known attacks, the other system, B2, did not show lower EER and min-tDCF because this LFCC-based system is already poor in detecting those attacks (see results for A04 and A06 in Table~\ref{Table:LA_CM_Baseline_Dev}). In comparison, we noticed that attacks A10, A13, and A17 had a higher EER and min-tDCF consistently for both the B1 and B2 systems. A10 uses the Tacotron acoustic model with WaveRNN for waveform generation, whereas A13 and A17 employ an acoustic model based on the moment matching neural network and VAE, respectively, with waveform filtering for waveform generation. The attacks included in the training do include similar methods for waveform generation. From the results of A17, we can also conclude that the acoustic model seemed to have less of an impact than the waveform generation method since spoofed audio data created with the VAE based acoustic model was already used in CM training and the waveform generation method is different from the methods used in training. This was also confirmed by the comparison of performance obtained with A10 and A11 where the only difference was in the waveform generation method. We noticed that attacks A08, A09, and A11 were easy to detect for both baseline CM systems. It turns out that the waveform generation methods based on the neural source-filter model, Vocaine, and Griffin-Lim were not able to produce  audio that shows more similarity to genuine speech under the current settings. Finally, when comparing the two CM methods, we found that B1 was better than B2 in most cases. However, for A09, A13, A14, and A18, the LFCC-based B2 system performed better than B1.

\begin{figure*}[t!]
\begin{center}
\includegraphics[clip,trim=4cm 0 {.5\wd0} 0,width=20cm]{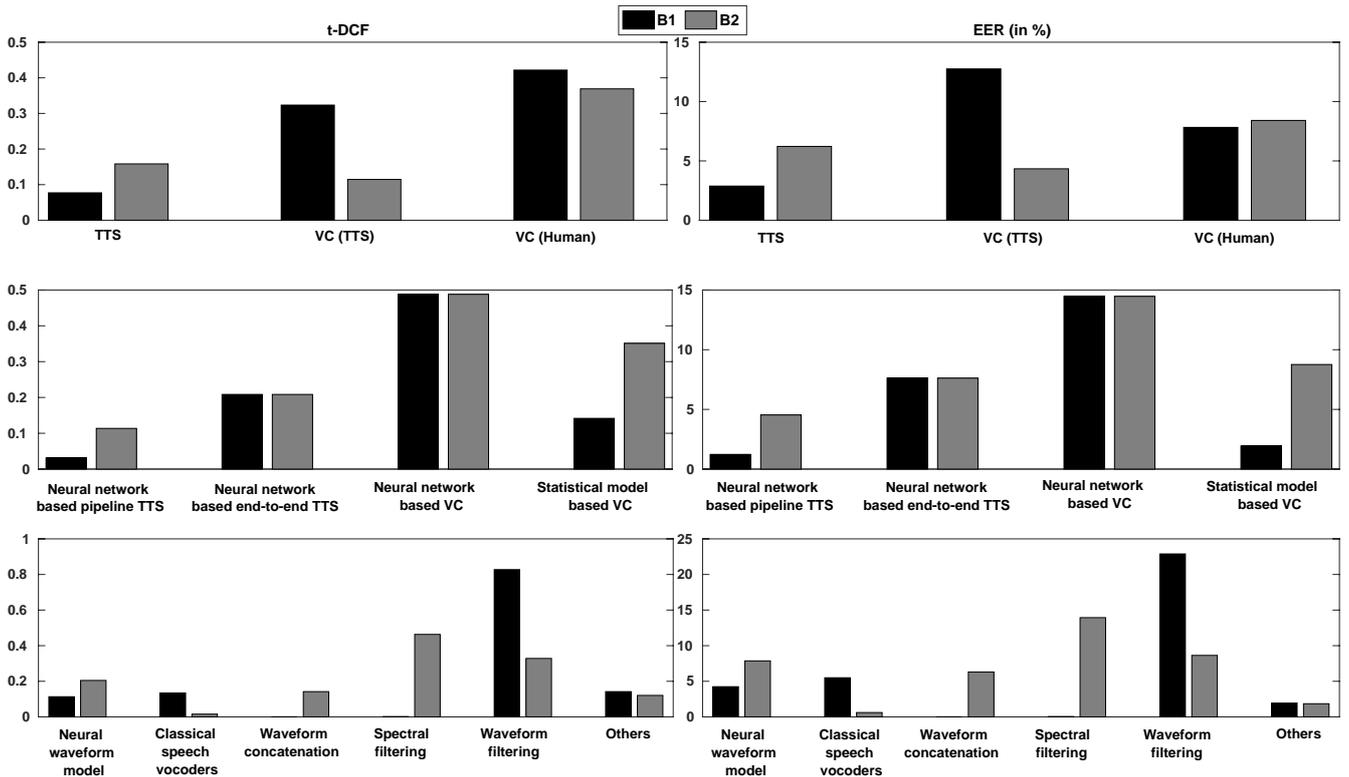}\\
\caption{Summary of LA subset (evaluation section) results in terms of t-DCF (left) and EER (right). First row shows results for three categories of synthetic speech: (i) TTS, (ii) VC (TTS), and (iii) VC (Human). Next row shows results for four types of acoustic models: (i) neural-network-based pipeline TTS, (ii) neural-network-based end-to-end TTS, (iii) neural-network-based VC, and (iv) statistical-model-based VC. Last row shows results for different waveform generation methods: (i) neural waveform models, (ii) classic speech vocoders, (iii) waveform concatenation, (iv) spectral filtering, (v) waveform filtering, and (vi) others.}
\label{Fig:DetailsLAEval}
\end{center}
\end{figure*}

In Figure~\ref{Fig:DetailsLAEval}, we have summarized the baseline results for the LA-evaluation subset according to the categories, acoustic models, and waveform generation methods. We computed the average performance over all the attacks for each type. We observed that TTS-based synthetic speech was easier to detect on average, especially when the B1 method was used. VC (Human) had the largest min-tDCF with the B1 method as the CM. Comparing the acoustic models, both the average min-tDCF and EER had lower values with neural-network-based pipeline TTS methods (A07, A08, A09, and A12). In contrast, attacks created with acoustic modeling using the neural-network-based VC methods (A13, A14, A15, A17) were the most difficult to detect with the highest average min-tDCF and EER. Finally, comparison of waveform generation methods demonstrates that synthetic speech created with waveform-filtering-based approaches (A13 and A17) were the most difficult to detect than the other methods.


\subsection{Analysis of PA subset}

\textcolor{black}{Reported here are results for the PA subset.  Two sets of results are presented in order to show the influence of different replay configurations and different acoustic environments upon baseline ASV and CM performance.  In contrast to those presented for the LA subset above, results for the PA subset are presented for the evaluation set only.  Differences between LA spoofing attacks in the development and evaluation partitions correspond to fundamentally different algorithms. There are then substantial differences between ASV and CM performance in the case of each attack algorithm.  For the PA scenario, differences between development and evaluation partitions correspond to differences only in specific impulse responses; acoustic environment and replay configuration categories are exactly the same.  In this case, differences in ASV and CM performance are far less substantial and do not warrant unnecessary attention here.  Results for the PA development set are entirely consistent with the trends reported for the evaluation set and are nonetheless available in the documentation that accompanies the ASVspoof 2019 database.}

\subsubsection{Replay configurations}

\begin{figure*}[t!]
\begin{center}
\includegraphics[clip,width=18cm]{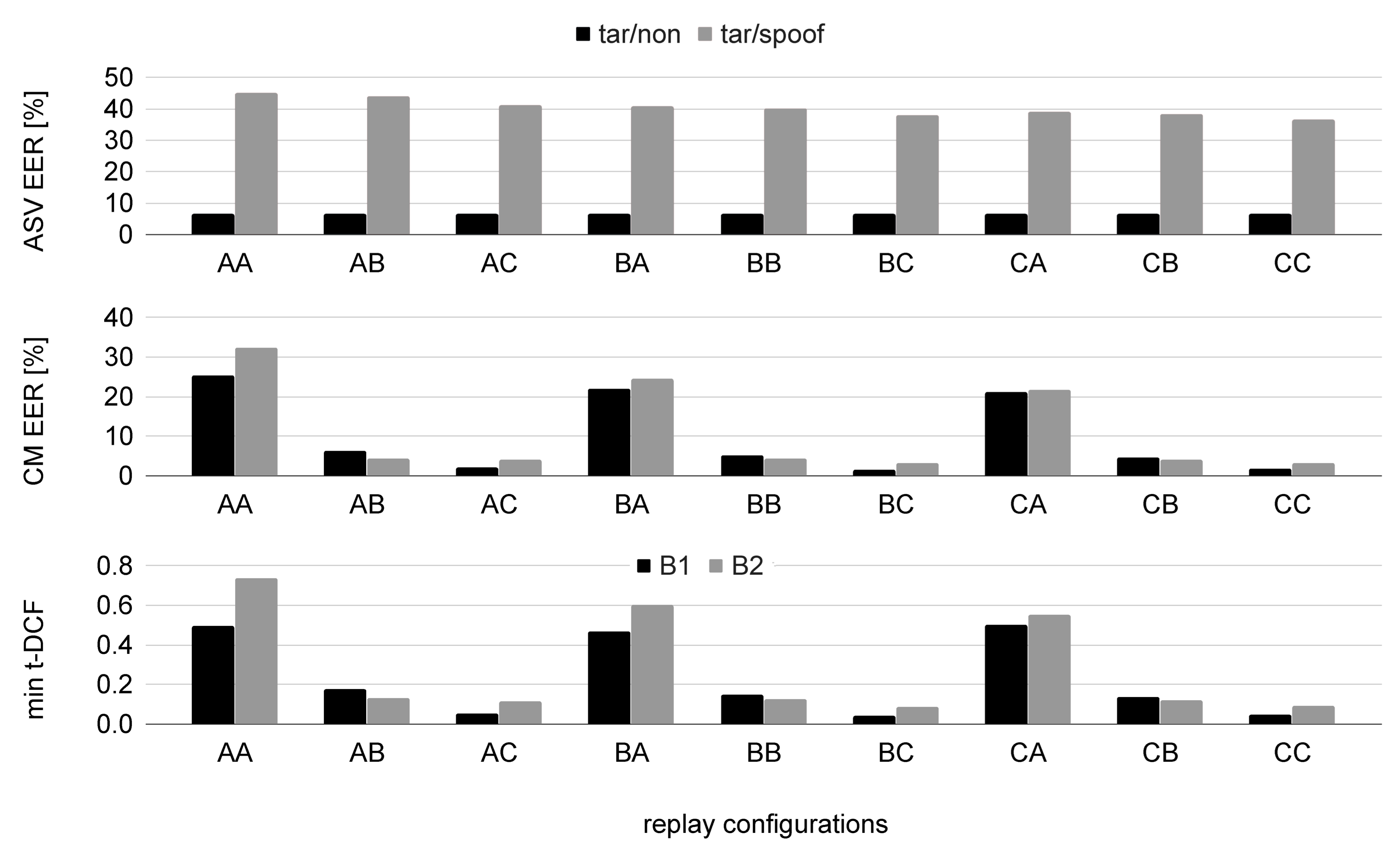}\\
\caption{An illustration of baseline results for the PA scenario of the ASVspoof 2019 database.  Results illustrated for individual replay configurations (pooled acoustic environments) and for: (top panel) standalone ASV results in terms of EER (\%) with target and zero-effort impostor trials (black bars) and target and replay spoofing trials (gray bars); (middle panel) standalone replay spoofing in terms of EER (\%) for baselines B1 and B2; (bottom panel) combined ASV and CM results illustrated in terms of the min-tDCF.}
\label{Fig:DetailsPAEval_replayconfig}
\end{center}
\end{figure*}

\textcolor{black}{ASV performance for the PA scenario is illustrated in the top panel of Figure~\ref{Fig:DetailsPAEval_replayconfig}.  The plot shows standalone ASV performance in the case of target and zero-effort impostor trials (black profile) and then target and replay spoofing trials (grey profile).  Results are shown separately for each of the nine different replay configurations, each denoted by the duple $\textup{D}_a$,Q (see Table~\ref{tab:spoofing_technique_att}).  In all cases, results are pooled across the full set of 27 different acoustic environments.  All results are reported in terms of the EER.}

\textcolor{black}{Baseline EERs are somewhat higher for PA than they are for LA. This is due to the session variability introduced in each acoustic environment, \emph{e.g.}, variability in convolutive channel noise.  EERs for replay spoofing conditions are markedly higher.  The highest EERs are observed for replay configurations with the shortest attacker-to-talker distance ($\textup{D}_a$) and the highest quality loudspeaker Q (condition AA).  The EER decreases monotonically as either the attacker-to-talker distance 
increases or the device quality decreases.  Among the different replay configurations, the lowest EER is observed for the largest attacker-to-talker distance and the lowest quality loudspeaker (condition CC).}



\textcolor{black}{Baseline countermeasure performance for the PA scenario is illustrated in the middle and bottom panels of Figure~\ref{Fig:DetailsPAEval_replayconfig} in terms of the EER and min-tDCF respectively.  In both cases, results are presented for the two baseline countermeasures B1 and B2.  Trends are consistently similar to those observed for standalone ASV performance.  The highest min-tDCF and EER results are for the shortest attacker-to-talker distances ($\textup{D}_a$) and the highest quality devices Q (condition AA); higher quality replay attacks are more difficult to detect. Both the min-tDCF and EER decrease when either the attacker-to-talker distance increases or the device quality decreases.  The lowest min-tDCF and EER results are obtained for the largest attacker-to-talker distances and the lowest device quality.}
\textcolor{black}{Also observed are the relative impacts of attacker-to-talker distance and the device quality.  The device quality has a considerably greater impact on baseline CM performance.  As is the case for the LA scenario, for the majority of cases, baseline B1 outperforms B2.}  

\begin{figure*}[t!]
\begin{center}
%
\includegraphics[clip,width=18cm]{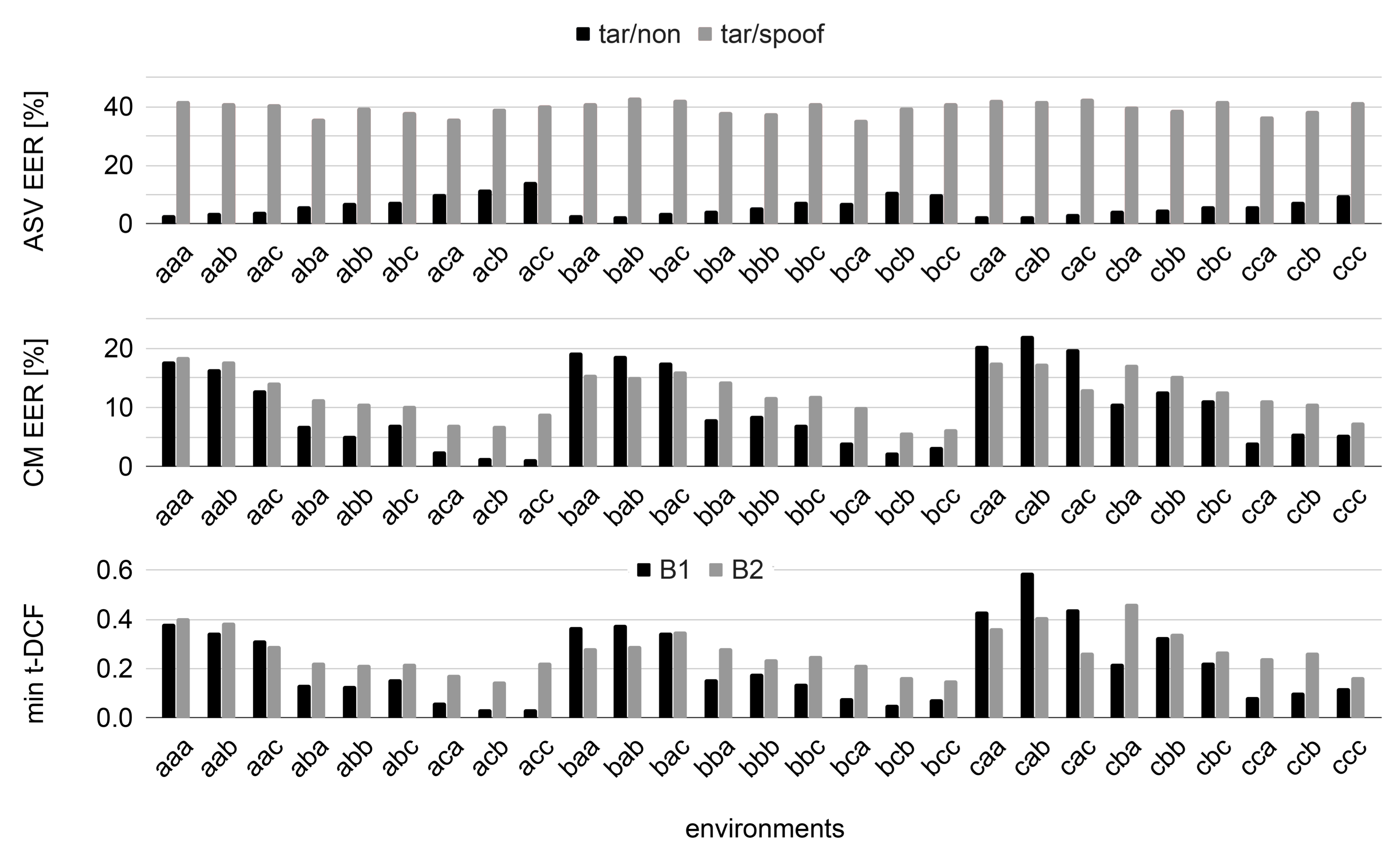}\\
\caption{\textcolor{black}{As for Figure~\ref{Fig:DetailsPAEval_replayconfig} except for results in terms of individual acoustic environments (pooled replay configurations).}}
\label{Fig:DetailsPAEval_acoustic}
\end{center}
\end{figure*}

\subsubsection{Acoustic environment}

\textcolor{black}{Figure~\ref{Fig:DetailsPAEval_acoustic} shows results in terms of the 27 different acoustic environments each denoted by the tuple S,R,$\textup{D}_s$ (see Table~\ref{tab:spoofing_technique_env}).  In a fashion consistent to results shown in Figure~\ref{Fig:DetailsPAEval_replayconfig} for replay configurations, they are shown for standalone ASV~(top panel), standalone CMs~(middle panel) and integrated ASV and CMs~(bottom panel).  Once again, standalone results are illustrated in terms of the EER, whereas integrated results are in terms of the min-tDCF.}

\textcolor{black}{Baseline ASV results (no spoofing) shown in the top panel of Figure~\ref{Fig:DetailsPAEval_acoustic} show that the room size S has little influence upon performance, whereas higher T60 reverberation times R and larger talker-to-ASV distances $\textup{D}_s$ result in higher EERs/min-tDCFs.  Replay spoofing results in substantial increases in the EER for all acoustic environments.}

\textcolor{black}{Baseline CM results expressed either in terms of the EER or min-tDCF show different trends.  While the room size still has little influence, \emph{higher} T60 reverberation times R and talker-to-ASV distances now give \emph{lower} EERs/min-tDCFs.  This observation is consistent with intuition.  Replay attacks act to amplify the effects of the acoustic environment and when these are greater, then replay detection can be performed with ease.  In contrast, when the acoustic environment is more benign, then replay detection is more challenging.}

\section{Human assessment}
\label{S:8}

While Section 6 demonstrated how the baseline ASV and CM performed on the ASVspoof 2019 database, we describe in this section a human assessment on the LA subset of the database. The results show whether the spoofed data was perceptually similar to the bona-fide data of target speakers and whether the spoofed data could be detected by the human subjects.
{Note that we did not conduct human assessment on the PA subset because the human perception of the bona fide and replayed speech may be affected by the varied types of headphones used by the human subjects during the crowd-sourced assessment. This uncontrolled acoustic environment will make it difficult to interpret the assessment results on PA subset.}

We designed the human assessment protocol on the basis of the ASVspoof LA evaluation protocol. The original protocols are unsuitable for human assessment because the prohibitive cost of evaluating the large number of spoofed trials in the evaluation set. In addition, the human subjects' perception may be biased by the unbalanced amount of spoofed and bona-fide evaluation data. 

For one target speaker in the evaluation set, we randomly selected 25 spoofed utterances from each of the 13 spoofing TTS/VC systems (A07 - A19). Meanwhile, we randomly sampled 25 bona-fide utterances from each of the 10 non-target speakers with the same gender as the target speaker\footnote{Because the number of non-target male speakers was 9, we over-sampled the data from one of the non-target male speakers.}. Since we collected 325 ($13 \times 25$) spoofed utterances and 250 ($10 \times 25$) non-target bona-fide utterances for the target speaker, to strike a balance between target and non-target data, we over-sampled 575 bona-fide utterances from the target speaker. We prepared the same amount of data for each of the 48 target speakers in the evaluation set and collected 27,600 ($48 \times 575$) bona-fide target utterances, 15,600 ($48 \times 325$) spoofed utterances, and 12,000 ($48 \times 250$) bona-fide non-target utterances, which lead to 55,200 utterances in total. 

During the assessment, the subjects were asked to conduct two role-playing tasks \cite{7400997} in one evaluation page. In one task, they listened to one utterance and were asked to judge whether the utterance was produced by a human or machine, given an imagined scenario where he or she must detect abnormal telephone calls in the customer service center of a commercial bank. The played utterance was randomly selected and may be bona-fide target, non-target, or spoofed. The subject was asked to give a score ranging from 1 to 10, where 1 indicates that the utterance was absolutely machine-generated, and 10 denotes a human-produced utterance with total confidence. The instruction given to subjects is as follows.
\begin{quotation}
Imagine you are working for a bank call center. Your task is to correctly accept only inquiries from human customers and to properly determine those that may be due to artificial intelligence as `suspicious cases that may be malicious'. However, if almost everything is judged to be `artificially generated speech', there will be many complaints from real customers, which must be avoided. Imagine a situation where you are working to protect bank accounts and balance convenience.

The audio sample that you will listen to is audio produced by humans or produced artificially by artificial intelligence. There are not only a system that sounds unnatural like a robot but also an artificial intelligence system that synthesizes natural speech that is very similar to human speech.

Now, please listen to the audio sample and determine whether the voice is artificially generated by artificial intelligence or is uttered by a person on the basis of only the voice you hear. You can listen to it as many times as you like. The content of the conversation in English is irrelevant and does not need to be heard. Please judge on the basis of only the characteristics of the sound, not the content of the words.
\end{quotation}

\begin{figure*}[!t]
	\centering
	\includegraphics[width=1.0\textwidth]{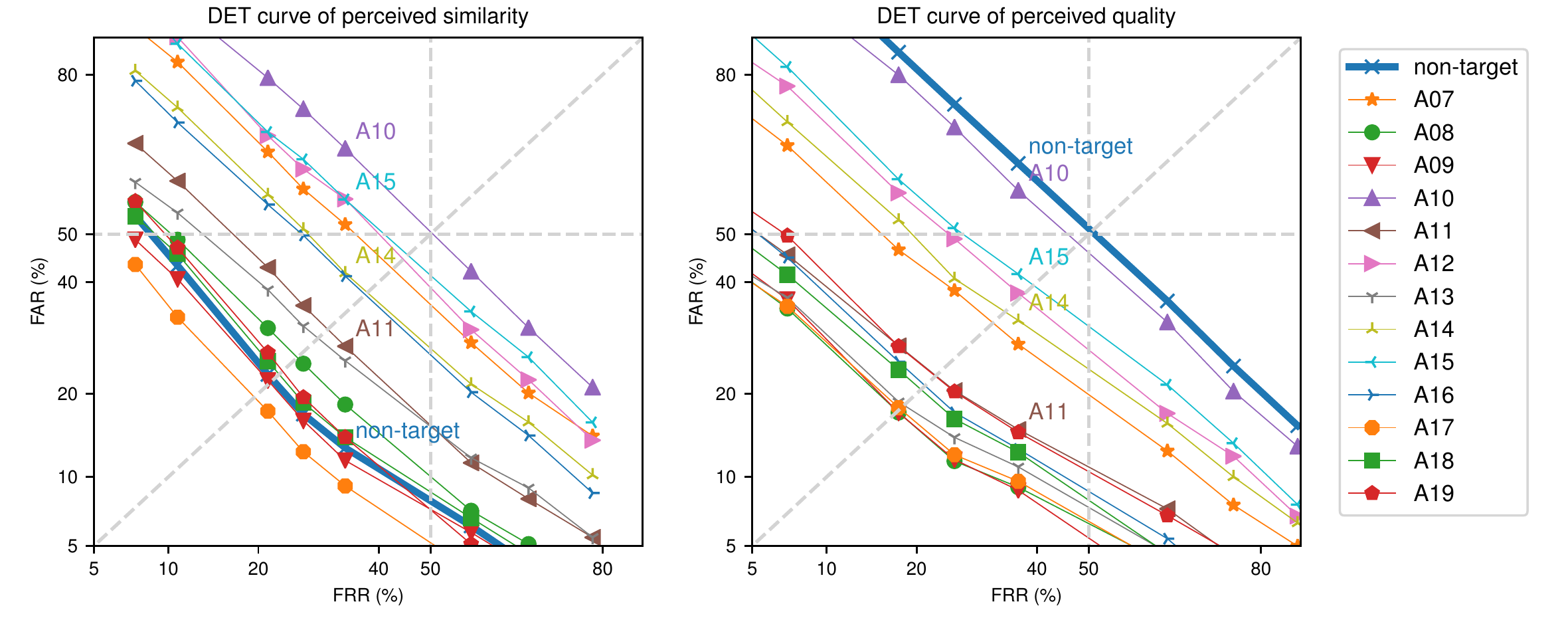}
	\caption{DET curves based on human assessment of similarity to target speakers (left) and speech quality (right)}
	\label{fig:human_similarity}
	\label{fig:human_quality}
\end{figure*}

In the other task, the subjects listened to two utterances and was asked to judge whether they sounded like the voice of the same person. Again, the subjects were instructed to imagine a scenario in which he or she must judge whether the voice in an incoming telephone call is similar to a recorded one that the speaker claimed. For this scenario, the recorded voice was a randomly selected enrollment utterance of the target speaker, and the stimuli may have been a bona-fide utterance or a spoofed utterance of that speaker or a bona-fide utterance of a non-target speaker. The subject was asked to evaluate the similarity using a scale of 1 to 10, where 1 and 10 denote `different speakers' and `the same speaker' with absolute confidence. The detailed instruction given to subjects is as follows:
\begin{quotation}
As before, imagine you are working for a bank call center. Your next task is to compare customer inquiries with voices recorded when the same customer made inquiries in the past. From the voices, you must determine whether the voices are of the same person or another person who is impersonating the original voices. However, if you choose `spoofing by someone else' more than necessary, there will be many complaints from real customers, which should be avoided. Imagine a situation in which you are working to protect bank accounts and balance convenience.
Now press the `Sample A' and `Sample B' buttons below and listen to the samples. You can listen to them as many times as you like. Use only the audio you hear to determine if the speakers are the same or not. The content of the conversation in English is irrelevant and does not need to be heard. Please judge on the basis of the characteristics of the voice, not the content of the words. If the sound is artificially generated, please judge it as a different speaker.
\end{quotation}

We prepared 55,200 evaluate pages that covered all the collected utterances and organized the human assessment on a crowd-sourcing platform. In total, 1,145 subjects participated, and each of them evaluated at least 32 pages. We acquired around 27,000 scores for bona-fide utterances of target speakers, around 12,000 scores for bona-fide utterances of non-target speakers, and around 1,200 scores for spoofed utterances produced by each of the TTS/VC spoofing systems in the CM evaluation set.

We first drew DET curves on the basis of  the similarity scores. The results are plotted in Figure~\ref{fig:human_similarity}.
On the basis of the scores of the other task, we drew DET curves where the bona-fide data from the target speakers was assumed to be positive and the bona-fide data from the non-target speakers and spoofed data negative.
The results are plotted in Figure~\ref{fig:human_quality}. 

As expected, we see that most of the spoofing systems had a DET curves above that of the non-target data in Figure~\ref{fig:human_similarity}, which indicates that these spoofing systems generated spoofed data that sounded similar to the target speakers in varied degrees. In Figure~\ref{fig:human_quality}, the DET curves of the spoofed data spread across the plane, suggesting a varied perceived quality and similarity in the spoofed data between the different spoofing systems. In general, we can see that the spoofed data in the LA database had disparate perceived quality and similarity to the target bona-fide data. 

Among the spoofing systems, the DET curve of A10 was close to the diagonal line in Figure~\ref{fig:human_similarity}. 
According to the results of a Mann-Whitney U test~\cite{rosenberg2017bias}, the difference between the spoofed data from A10 and the bona-fide utterances of the target speakers was marginally significant ($p=0.012$) in terms of the quality and insignificant ($p=0.81$) in terms of the similarity to the enrollment utterances. 
These results demonstrate that the spoofed utterances from A10 sounded the same as the target bona-fide speaker and sounded very natural from the human listener's perspectives, clearly showing that the state-of-the-art TTS has the capability of producing synthetic speech that is perceptually indistinguishable from bona fide speech. 
Since A10 also lead to a high EER for both the ASV and CM baselines, we can summarize this as A10 being good at fooling both humans and the ASV/CM baselines. 


We can also see that the spoofed data from A11 was much more easily detected in the human assessment, even though A11 used the same TTS architecture as A10, except that A11 and A10 used Griffin-Lim and WaveRNN for waveform generation, respectively. Similar results can be observed by comparing A14 and A15, which were identical in the TTS-VC architecture but differed in terms of waveform generation. Both the difference between A10 and A11 and that between A14 and A15 were statistically significant ($p<0.001$).
These results suggest that the perceived quality of spoofed data is affected by the waveform generation technique used. Furthermore, the perceived similarity to the target speaker is also influenced by the performance of the waveform generator. 

Interestingly, the human assessment also showed different results from those obtained using the baseline ASV and CM. For example, while A13 was good at fooling the baseline ASV and CM, it was easily detected by the human evaluators as spoofed data as Figure~\ref{fig:human_quality} shows. Likewise, A17 was also good at fooling the baseline CM, but it was also easily detected by the human evaluators. 

In all, the LA subset of the ASVspoof2019 database includes spoofed data with varied degrees of perceived quality and similarity to the target speakers. Some of them are strong spoofing methods even for the baseline ASV and CM. Such variation is essential for the LA task so that the performance of ASV and CM can be examined under various conditions.

\section{Summary}
\label{S:9}

This paper described the design, protocol, and spoofing attack implementations of the ASVspoof 2019 database. This is the first database that considers all three types of spoofing attacks, and it has two specific use case scenarios: logical access and physical access scenarios. Spoofing attacks within the logical access scenario have been generated with the latest speech synthesis and voice conversion technologies, including state-of-the-art neural acoustic and waveform model techniques, and have varied degrees of perceived quality and similarity to the target speakers, including spoofed data that cannot be differentiated from bona fide utterances even by human subjects. Replay spoofing attacks within the physical access scenario have been generated through carefully controlled simulations that support much more revealing analysis than possible previously. Both the acoustic environment and attacks' factors, including the physical placement of the microphones and loudspeaker in the acoustic environment, were systematically simulated. This paper also reported their impact on the reliability of the baseline ASV systems and two spoofing countermeasures and their performance using the traditional metric EER and newly introduced metric t-DCF for each scenario. 

This database, including ground truth labels, meta labels, and baseline ASV scores, is freely available for both academic and commercial purposes at the Edinburgh DataShare\footnote{\url{https://doi.org/10.7488/ds/2555}} under the Open Data Commons Attribution License. We strongly believe that the ASVspoof 2019 database will further accelerate and foster research on speaker recognition and media forensics. 

\section{Acknowledgements} 
The work was partially supported by JST CREST Grant No.\ JPMJCR18A6 (VoicePersonae project), Japan, MEXT KAKENHI Grant Nos.\ (16H06302, 16K16096, 17H04687, 18H04120, 18H04112, 18KT0051), Japan, the VoicePersonae and RESPECT projects funded by the French Agence Nationale de la Recherche~(ANR), 
the Academy of Finland (NOTCH project\ no.\ 309629 entitled ``NOTCH: NOn-cooperaTive speaker CHaracterization''). The authors at the University of Eastern Finland also gratefully acknowledge the use of the computational infrastructures at CSC -- the IT Center for Science, and the support of the NVIDIA Corporation the donation of a Titan V GPU used in this research. The numerical calculations of some of the spoofed data were carried out on the TSUBAME3.0 supercomputer at the Tokyo Institute of Technology. The work is also partially supported by Region Grand Est, France. The ADAPT centre (13/RC/2106) is funded by the Science Foundation Ireland (SFI).

\bibliographystyle{elsarticle-num}
\bibliography{main.bib}







\end{document}